\documentclass[11pt]{article}
\usepackage{graphicx}
\usepackage{epstopdf}

\newcommand{\be}{\begin{eqnarray}}
\newcommand{\ee}{\end{eqnarray}}

\def\ll#1{\left#1}
\def\r#1{\right#1}
\def\fr{\frac{1}{2}}

\def\mref#1{(\ref{#1})}

\def\p{\partial}

\def\bd{\begin{displaymath}}
\def\ed{\end{displaymath}}

\def\ba#1{\begin{array}{#1}}
\def\ea{\end{array}}
\def\nn{\nonumber}
\newfont{\Bbb}{msbm10 scaled 1200}

\newtheorem{tw}{THEOREM}
\newtheorem{de}{DEFINITION}

\begin{document}

\pagestyle{empty}

\begin{center}

{\LARGE\bf The semiclassical states excitations in the multi-rectangular billiards \\[0.5cm]}

\vskip 12pt

{\large {\bf Stefan Giller}}

\vskip 3pt

Jan D{\l}ugosz University in Czestochowa\\
Department of Experimental and Applied Physics\\
Armii Krajowej 13/15, 42-200 Czestochowa, Poland\\
e-mail: stefan.giller@ujd.edu.pl
\end{center}

\vspace {10pt}

\begin{abstract}
The problem of the quantizations of the $L$-shaped billiards and the like ones, i.e. each angle of which is equal to $\pi/2$ or $3\pi/2$,
is considered using as a tool the Fourier series expansion method. The respective wave functions and the quantization
conditions are written and discussed looking for and discussing about the superscars effects in such multi-rectangular billiards (MRB). It is
found that a special set of POC
modes effect the superscars phenomena in MRB in which the billiards are excited as a whole to the modes closest to the
semiclassical ones existing in their approximated copies being MRB in which their parallel sides remain in rational relations between themselves.
\end{abstract}

\vskip 10pt
\begin{tabular}{l}
{\small PACS number(s): 03.65.-w, 03.65.Sq, 02.30.Jr, 02.30.Lt, 02.30.Mv} \\[1mm]
{\small elementary polygon patterns, Riemann surfaces, periodic orbit channels, superscar states}\\[1mm]
\end{tabular}

\newpage

\pagestyle{plain}

\setcounter{page}{1}

\section{Introduction}

\hskip+2em Since the discovery by Richens and Berry \cite{8} that the rational polygon billiards (RPB) with the exception of the small number of
them are pseudointegrable, i.e. not chaotic, the billiards have attracted much attention to answer the questions about properties of their
respective energy spectra and wave functions. The methods of finding the exact answers for these
questions were mainly numerical \cite{8,21,22,2,15} while the approximating ones have been focused mainly on using the Gutzwiller semiclassical trace formula
\cite{19}. Other
rather rare attempts of considering the problem to get some explicit formulae for both the energy spectra and the respective wave functions also
have used the semiclassical limits \cite{23,3}. It is striking however that in these investigations except the last papers mentioned the
rationality of RPB has not been used explicitly to get the respective
results, i.e. a direct relation between the rationality of RPB and the properties of the wave functions and energy spectra in RPB has not been
established.

However as it was said above the rationality of RPB has been used directly and explicitly in the papers \cite{23,3} but only in the semiclassical
limit which the method has dropped unfortunately the majority of the wave functions and the spectra of the considered RPB. Nevertheless the
respective semiclassical considerations have been possible by using the idea of doubly rational polygon billiards (DRPB).

In the series of papers
of Bogomolny and of Bogomolny {\it\underline{et} \underline{al}} \cite{6,16,18} the rationality of RPB expressed by the existence in them of many periodic
orbit channels (POC) was used to analyse at the
numerical level the relation (superscars effects \cite{10}) between the (simple) energy spectra of POC
present in RPB considered and both its exact spectrum and the wave functions which the relation has shown the closeness
between the respective energy spectra leaving however its origin not well established. Some trials for that were done however in our earlier
paper \cite{3} but at the semiclassical limit only showing that the energy spectra of some POC can be parts of the spectra of the RPB considered.

As challenging problems for the explicit quantizations of RPB are considered their vertices being the singular points of their boundaries.
From the quantum mechanical point
of view the vertices are sources of so called strong diffractions for the billiards wave functions providing most troubles in handling them
analytically \cite{1}. Nevertheless some quite general limitations for respective properties of the wave functions in RB close to their vertices
have been established \cite{12}. However still new approach to the explicit quantizations of RPB was developed by our another paper \cite{11} where
the idea
of Riemann surfaces was formulated and realized for a class of RPB called POC developed RPB (POCDRPB). In developing the idea the vertices
appeared to
be just the branch points on the respective surfaces and the wave functions defined in POCDRPB when continued on the respective Riemann surface
appeared to be periodic functions on it allowing to apply for their analysis the Fourier series methods in which the series appeared to be
nothing but the expansions of the wave functions by the eigenfunctions of the respective POC building the surface.

Just the idea of the Riemann surface when confined in the present paper to the case of POCDRPB the sides of which were orthogonal to each other
in every of its vertex, i.e. to MRB, was possible to be greatly simplified avoiding the
constructions of the Riemann surfaces themselves but still allowing for direct applications of the Fourier series method. This simplification  was possible due
to the fact that for any MRB its so called elementary polygon pattern (EPP) - the basic element periodically constructing the respective Riemann surface - is
composed only of the four mirror images of MRB. As a result one got relatively simple forms of the stationary wave functions together with the respective conditions
for their energy levels.

We would like to stress that the Fourier series method used in the paper to expand the wave functions are the pure ones, i.e. they are handbook
series \cite{13} taking into account some their essential properties described in App.A of our paper. This differ essentially
our paper from the ones of Richens and Berry \cite{8} and of Wiersig \cite{2,15} who used in their papers rather unusual series with unknown
properties to compute energy levels and wave functions for RPB they considered.

The method used allowed us further for a detailed analyses of the wave functions found looking for the superscars effects in some of them.
Surprisingly the result found was the existence of at least two kind od superscars states - the one generated by single
POC present in the billiards considered and the other in which states of several POC resonate simultaneously being formed of semiclassical modes of
MRB rationally approximating the
original one, i.e. of the semiclassical modes of the doubly rational MRB (DRMRB). Just the latter superscars states were possible to be studied by the
Fourier series method used in the paper while the former still need rather more sophisticated approach to confirm theoretically their existence.
These superscars and the single POC exciting them are those which were investigated numerically and experimentally by Bogomolny and Bogomolny
{\it\underline{et} \underline{al}} \cite{6,16,18}.

The paper is organized as follows.

In Sec.2 it is shown how the construction of the $L$-shaped billiards (LSB) wave functions helping by the respective Riemann surface can be simplified and reduced to LSB itself.

In Sec.3 the detailed quantization of LSB  by the Fourier series expansions is performed providing the respective wave functions
and the quantization conditions for the energy levels. In the same section the method is generalized to any multi-rectangular billiards containing
also multi-rectangular holes inside.

In Sec.4 the quantization conditions established for LSB are used to analyse their possibility for permitting some energy levels
determined by them to be close as much as possible to the ones of the horizontal and vertical POC existing in the billiards.

First it is shown that our method divides modes of the horizontal and vertical POC which can be excited in the billiards into two classes - the
one which collects modes called semiclassical and which are studied in the paper and the remaining modes which possible existence and properties are
not established the paper.

The modes of the first class appear to belong to the semiclassical modes of DRLSB approximating the original LSB and
which can resonate dominating the respective modes of LSB by single terms of the respective Fourier series which coincide simultaneously with
the modes of each of their both POC. In particular it is shown that the superscar states can be exited in an arbitrary LSB being close to the
semiclassical ones of the respective DRLSB approximating the former with the controlled accuracies. There are infinitely
many of such superscar states resonating in the original LSB correspondingly to its infinitely many DRLSB copies
approximating it still closer and closer. Their existence was confirmed experimentally by Kudrolli and Sridhar \cite{20}.

The essential difference between the two kinds of the superscars modes lies in the possibility of the first kind modes to be controlled by the
basic theorems on the proximity of two spectra corresponding to two slightly different areas (see App.B) while for the second kind of the superscars
modes such a control cannot be applied since as a rule a single POC cannot cover the area of so called elementary polygon pattern (EPP) \cite{23}
prescribed to each RB except for the trivial case of the rectangular billiards. The latter property is essential for applying the theorems of App.B.

The result established in this section for LSB is valid also for any multi-rectangular billiards which is shown in the same section.

The paper is finished by Sec.5 summarizing and discussing its results.

\section{Getting wave functions in LSB by folding the ones defined on the LSB Riemann surface}

\hskip+2em Let us remind the construction of the Riemann surface (RS) corresponding to LSB. The latter is illustrated by Fig.1.
\begin{figure}
\begin{center}
\includegraphics[width=15cm]{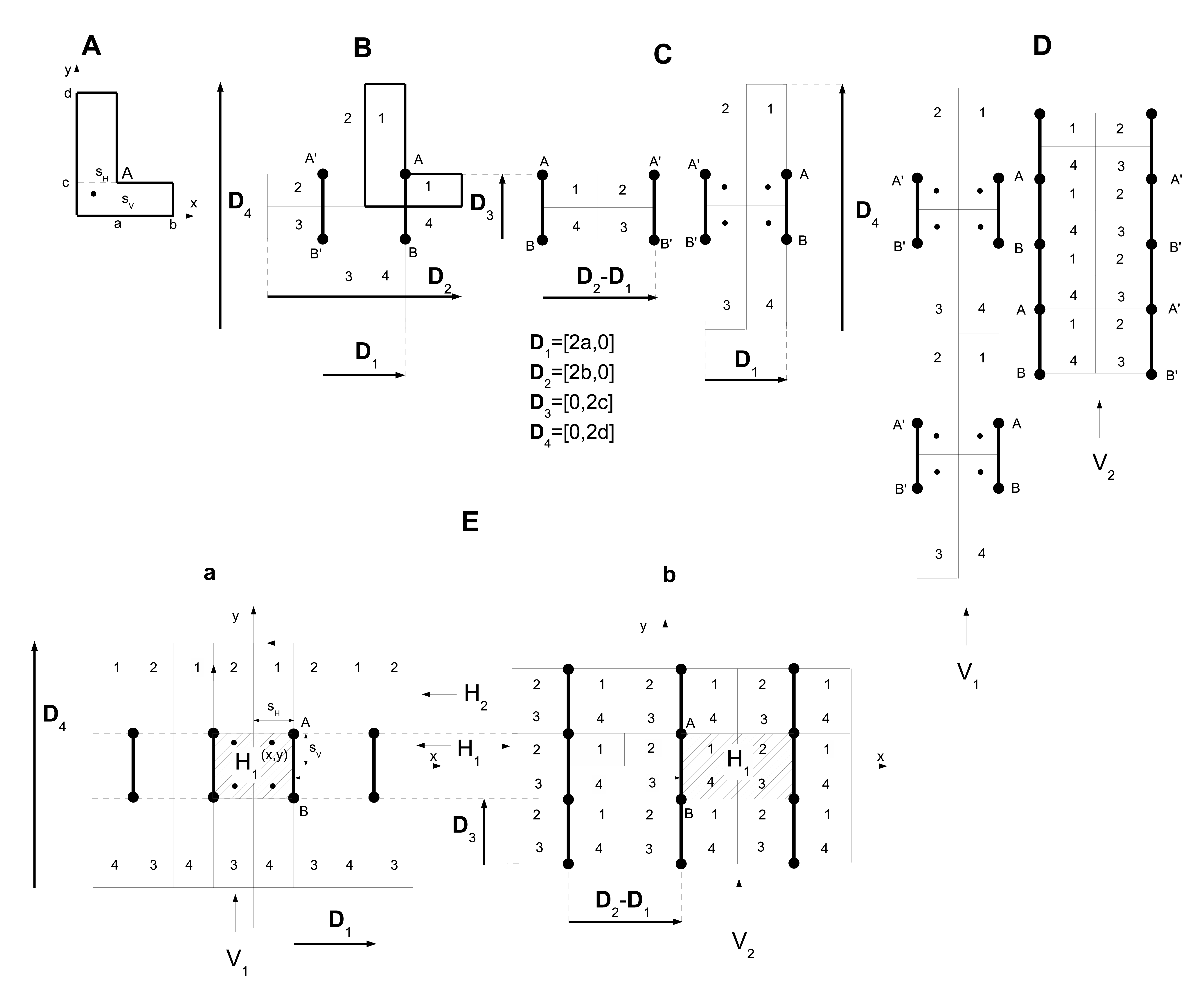}
\caption{The Riemann surface construction for the $L$-shaped billiards \cite{11}.  {\bf A} - the billiards; {\bf B} - its corresponding elementary polygon pattern (EPP)  cut along the two singular diagonals $A-B, \;A'-B'$; {\bf C} - the two basic elements forming POC ${\bf V}_1$ and ${\bf V}_2$ on Fig.2D; {\bf E} - the two sheets {\bf a} and {\bf b} each glued of the respective POC ${\bf V}_1$ and ${\bf V}_2$ and next glued with each other by a single cut  A-B - there are infinitely many of them which the Riemann surface is glued of}
\end{center}
\end{figure}
The so called presolution $\Psi(x,y)$ to SE is defined by two its branches given by their following Fourier series on RS
\be
\Psi_{V_1}(x,y)=\sum_{i,j=1,2}\sum_{m,n\geq 1}V_{mn}^{(1)}f_i\ll(m\pi\frac{x}{a}\r)f_j\ll(n\pi\frac{y}{d}\r)\nn\\
\Psi_{V_2}(x,y)=\sum_{i,j=1,2}\sum_{m,n\geq 1}V_{mn}^{(2)}f_i\ll(m\pi\frac{x}{b-a}\r)f_j\ll(n\pi\frac{y}{c}\r)
\label{1}
\ee
 where $f_1(x)=\sin(x)$ and $f_2(x)=\cos(x)$.

Note that the series in \mref{1} are nothing but the expansions of $\Psi(x,y)$ by the eigenfunctions in POC ${\bf V}_1$ and ${\bf V}_2$ when the latter are quantized with any pair of the boundary conditions on their sides.

The above branches should be matched with each other on the cuts by which the sheets of RS are glued with themselves. However it is not possible directly by using the series \mref{1} since both the branches are discontinuous on the cuts and are not given there by the series. Therefore to make the matching it is necessary to make use of another representation for $\Psi(x,y)$ by its Fourier series developed on the horizontal POC ${\bf H}_1$ crossing both POC ${\bf V}_1$ and ${\bf V}_2$ in the hatched areas in Fig.Da,b. The respective series is the following
\be
\Psi_{H_1}(x,y)=\sum_{m,n\geq 1}H_{mn}^{(1)}f_i\ll(m\pi\frac{x}{b}\r)f_j\ll(n\pi\frac{y}{c}\r)
\label{1b}
\ee

The further procedure depends now on the boundary conditions put on the solution $\Psi_L(x,y)$ in LSB on its sides. In general $\Psi_L(x,y)$ is formed by $\Psi(x,y)$ by the following folding formula \cite{11}
\be
\Psi_L(x,y)=\Psi_{V_1}(x,y)\pm\Psi_{V_1}(-x,y)\pm\Psi_{V_1}(-x,-y)\pm\Psi_{V_1}(x,-y)=\nn\\
\Psi_{H_1}(x,y)\pm\Psi_{H_1}(-x,y)\pm\Psi_{H_1}(-x,-y)\pm\Psi_{H_1}(x,-y)
\label{1c}
\ee
for the point $(x,y)$ in Fig.2Da.

A choice of signs in \mref{1c} depends on boundary conditions which not every set of them is possible to be used in the method developed in \cite{11}. For the Dirichlet ones however \mref{1c} gives for $\Psi_L(x,y)$ in the area $s_H\times s_V$ of Fig.2Da the following representation
\be
\Psi_L(x,y)=\Psi_{V_1}(x,y)-\Psi_{V_1}(-x,y)+\Psi_{V_1}(-x,-y)-\Psi_{V_1}(x,-y)=\nn\\
\sum_{m,n\geq 1}V_{mn}^{(1)}\sin\ll(m\pi\frac{x}{a}\r)\sin\ll(n\pi\frac{y}{d}\r)=
\sum_{m,n\geq 1}H_{mn}^{(1)}\sin\ll(m\pi\frac{x}{b}\r)\sin\ll(n\pi\frac{y}{c}\r)\nn\\
(x,y)\in s_H\times s_V
\label{1d}
\ee
in which the last equation matches the coefficients $V_{mn}^{(1))}$ with $H_{mn}^{(1)}$.

Now one can observe that the procedure applied above can be greatly simplified noticing that the boundary conditions leading to the final form \mref{1d} of $\Psi_L(x,y)$ can be put at the very beginning on the branches $\Psi_{V_1}(x,y)$ and $\Psi_{H_1}(x,y)$ demanding them
\begin{itemize}
\item to vanish on the segments of the sides of LSB traced on the RS sheets;
\item to be antisymmetric with respect to their central horizontal and vertical lines which the assumption is in agreement with previous one.
\end{itemize}
Note that both the assumptions do not determine in any way other properties of $\Psi(x,y)$ in LSB itself.

Making these assumptions we immediately get instead of the expansions \mref{1}-\mref{1b} the following ones
\be
\Psi_{V_1}(x,y)=\sum_{m,n\geq 1}V_{mn}^{(1)}\sin\ll(m\pi\frac{x}{a}\r)\sin\ll(n\pi\frac{y}{d}\r),\;\;\;\:(x,y)\in {\bf a}\nn\\
\Psi_{H_1}(x,y)=\sum_{m,n\geq 1}H_{mn}^{(1)}\sin\ll(m\pi\frac{x}{b}\r)\sin\ll(n\pi\frac{y}{c}\r),\;\;\;\:(x,y)\in {\bf b}
\label{1f}
\ee
which coincide with the ones for $\Psi_L(x,y)$ when the latter is expanded into the Fourier series in the respective  vertical or horizontal arms of LSB with the same Dirichlet boundary conditions put on $\Psi_L(x,y)$ on the sides of the arms.

Therefore the last observation allows us to avoid the whole procedure of construction of $\Psi_L(x,y)$ helping by RS and to start from the very beginning with the expansions \mref{1} performed directly in LSB rather then on RS considering $\Psi_{V_1}(x,y)$ and $\Psi_{H_1}(x,y)$ in \mref{1} as the branches of $\Psi_L(x,y)$ in LSB, see Fig.2. The vanishing of the expansions not only on the sides of LSB but also on the respective segments $s_V$ and $s_H$ in Fig.2 is typical for the Fourier series which do not reconstruct the expanded functions in their discontinuity points. Therefore $\Psi_{V_1}(x,y)$ and $\Psi_{H_1}(x,y)$ considered now on LSB cannot be matched on the segments $s_V$ and $s_H$ because of the same reason we noticed in the general procedure described above. So that the condition \mref{1d} of their coincidence in the rectangle $s_H\times s_V$ of LSB this time is still valid. How to gather with the expansions at their discontinuity points in LSB will be discussed in the next section.

\section{The multi-rectangular billiards and their quantization by the Fourier series}

\subsection{The $L$-shaped billiards}

\hskip+2em Let us start again with the simplest MRB, i.e. the $L$-shaped one shown in Fig.2 for which it is assumed from the very beginning that the ratios $c/d$ and $a/b$ are irrational.
\begin{figure}
\begin{center}
\includegraphics[width=10cm]{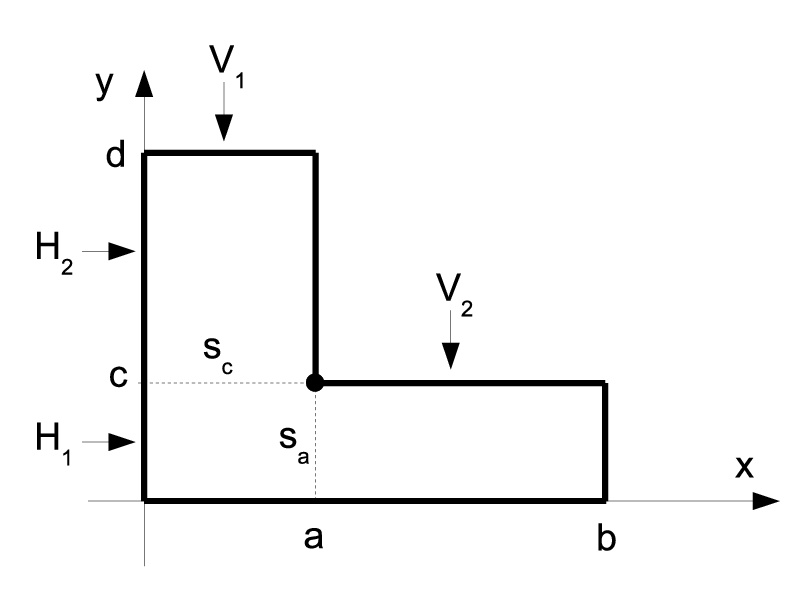}
\caption{The $L$-shaped billiards. The vertex distinguished on it by
the bold dot is the threefold branch point on the respective Riemann surface. The
horizontal and vertical broken lines emerging from the vertex are the singular diagonals for the respective POC $H_i$ and $V_i,\;i=1,2,$ \cite{6,11}}
\end{center}
\end{figure}

\subsubsection{The Dirichlet boundary conditions on all the sides}

\hskip+2em Let us assume again that the stationary wave function $\Psi(x,y)\equiv\Psi_L(x,y)$ in LSB we are looking for satisfies the Dirichlet boundary conditions on all the
sides of the billiards. According to the discussion in the previous section (see also  App.A ) $\Psi(x,y)$ has the forms \mref{1f} in the respective rectangles $b\times c$ and $a\times d$
so that
\be
\Psi_{H_1}(x,c)=\fr(\Psi(x,c)+(-\Psi(x,c)))=0,\;\;\;0<x<a\nn\\
\Psi_{V_1}(a,y)=\fr(\Psi(a,y)+(-\Psi(a,y)))=0,\;\;\;0<y<c
\label{1a}
\ee
i.e. $\Psi(x,c)$ is not determined on the segment $s_H$ by the series $\Psi_{H_1}(x,y)$ being however determined there by $\Psi_{V_1}(x,y)$ and
similarly $\Psi(a,y)$ is not determined by the series $\Psi_{V_1}(x,y)$ on the segment $s_V$ but instead it is determined by the series
$\Psi_{H_1}(x,y)$ there which the notes will be utilized below.

As it was discussed earlier the series  \mref{1f} representing the same function $\Psi(x,y)$ should be now matched in the rectangle $s_H\times s_V$ rather then on the segments $s_V$ and $s_H$ just because of the relations \mref{1a}. In fact both $\Psi_{H_1}(x,y)$ and $\Psi_{V_1}(x,y)$ should coincide there up to the second order of their derivatives since $\Psi(x,y)$ is demanded to be the $C^2$-class function in LSB. To write the equations confirming these identifications let
us calculate first the Fourier series coefficients for the derivatives of $\Psi(x,y)$ noticing that they cannot be got simply by the differentiations of the
series \mref{1f} because of the discontinuities of $\Psi_{H_1}(x,y)$ and $\Psi_{V_1}(x,y)$ on the respective segments $s_H$ and $s_V$.
Denoting by $X^{(x)},X^{(y)},X^{(xy)},X^{(x^2)},X^{(y^2)},\;X=H,V,$ the coefficients corresponding to the respective derivatives we have
instead (see App.A)
\be
\ba{ll}
H_{mn}^{(x)}=\frac{\pi m}{b}H_{mn}^{(1)}&V_{mn}^{(x)}=\frac{2(-1)^m}{a}v_n+\frac{\pi m}{a}V_{mn}^{(1)}\\
H_{mn}^{(y)}=\frac{2(-1)^n}{c}h_m+\frac{\pi n}{c}H_{mn}^{(1)}&V_{mn}^{(y)}=\frac{\pi n}{d}V_{mn}^{(1)}\\
H_{mn}^{(xy)}=\frac{\pi m}{b}H_{mn}^{(y)}&V_{mn}^{(xy)}=\frac{\pi n}{d}V_{mn}^{(x)}\\
H_{mn}^{(x^2)}=-\frac{\pi^2 m^2}{b^2}H_{mn}^{(1)}&V_{mn}^{(x^2)}=-\frac{2(-1)^m\pi m}{a^2}v_n-\frac{\pi^2 m^2}{a^2}V_{mn}^{(1)}\\
H_{mn}^{(y^2)}=-\frac{2(-1)^n\pi n}{c^2}h_m-\frac{\pi^2 n^2}{c^2}H_{mn}^{(1)}&V_{mn}^{(y^2)}=-\frac{\pi^2 n^2}{d^2}V_{mn}^{(1)}
\ea
\label{2b}
\ee
where
\be
h_m=\frac{2}{b}\int_0^a\Psi(x,c)\sin\ll(m\pi\frac{x}{b}\r)dx,\;\;\;
v_n=\frac{2}{d}\int_0^c\Psi(a,y)\sin\ll(n\pi\frac{y}{d}\r)dy,\;\;\;m,n\geq 1
\label{2}
\ee

Therefore the coincidence of $\Psi_{H_1}(x,y)$ and $\Psi_{V_1}(x,y)$ in the rectangle $a\times c$ takes the forms
\be
c\sum_{r\geq 1}\beta_{mr}H_{rn}^{(1)}=a\sum_{r\geq 1}V_{mr}^{(1)}\alpha_{rn},\;\;\;m,n\geq 1
\label{2a}
\ee
and
\be
c\sum_{r\geq 1}\beta_{mr}^{(i)}H_{rn}^{(i)}=a\sum_{r\geq 1}V_{mr}^{(i)}\alpha_{rn}^{(i)}\nn\\
i=x,y,xy,x^2,y^2
\label{2c}
\ee
where
\be
\alpha_{rn}=\int_0^c\sin\ll(r\pi\frac{y}{d}\r)\sin\ll(n\pi\frac{y}{c}\r)dy=
(-1)^n\frac{n}{\pi c}\frac{\sin\ll(r\pi\frac{c}{d}\r)}{\frac{r^2}{d^2}-\frac{n^2}{c^2}}\nn\\
\beta_{mr}=\int_0^a\sin\ll(r\pi\frac{x}{b}\r)\sin\ll(m\pi\frac{x}{a}\r)dx=
(-1)^m\frac{m}{\pi a}\frac{\sin\ll(r\pi\frac{a}{b}\r)}{\frac{r^2}{b^2}-\frac{m^2}{a^2}}\nn\\
\alpha_{rn}^{(y)}=\alpha_{rn}^{(xy)}=\frac{cr}{dn}\alpha_{rn}\nn\\
\alpha_{rn}^{(x)}=\alpha_{rn}^{(x^2)}=\alpha_{rn}^{(y^2)}=\alpha_{rn}\nn\\
\beta_{mr}^{(x)}=\beta_{mr}^{(xy)}=\frac{ar}{bm}\beta_{mr}\nn\\
\beta_{mr}^{(y)}=\beta_{mr}^{(x^2)}=\beta_{mr}^{(y^2)}=\beta_{mr}
\label{3}
\ee
while the coefficients $h_m$ and $v_n$ can be given the forms
\be
h_m=\frac{2}{b}\sum_{l,k\geq 1}\beta_{km}\sin\ll(l\pi\frac{c}{d}\r)V_{kl}^{(1)}\nn\\
v_n=\frac{2}{d}\sum_{l,k\geq 1}\alpha_{nl}\sin\ll(k\pi\frac{a}{b}\r)H_{kl}^{(1)}
\label{4}
\ee
when $\Psi(a,y)$ and $\Psi(x,c)$ are substituted in \mref{2} by their respective Fourier series $\Psi_{H_1}(a,y)$ and $\Psi_{V_1}(x,c)$.

However as it was established in App.A the matching conditions \mref{2c} can be reduced to the conditions \mref{2a}, i.e. the latter remain as
the unique ones.

Therefore the relations \mref{2a} realize the constructions of $\Psi(x,y)$ which vanishes on all the sides of LSB still
however not satisfying the Schr{\"o}dinger equation (SE).

Let us therefore join to them the latter equation making $\Psi(x,y)$ satisfying SE with an energy $\kappa^2=2E$, i.e. $\Psi(x,y)\to\Psi(x,y;\kappa)$.
In terms of the coefficients \mref{2b} the Schr{\"o}dinger equation can be rewritten as
\be
X_{mn}^{(x^2)}+X_{mn}^{(y^2)}+\kappa^2X_{mn}^{(1)}=0,\;\;\;X=H,V
\label{5}
\ee
or
\be
H_{mn}^{(1)}(\kappa^2-\kappa_{mn;bc}^2)=(-1)^n\frac{2\pi n}{c^2}h_m\nn\\
V_{mn}^{(1)}(\kappa^2-\kappa_{mn;ad}^2)=(-1)^m\frac{2\pi m}{a^2}v_n
\label{6}
\ee
where
\be
\kappa_{mn;bc}^2=\frac{\pi^2 m^2}{b^2}+\frac{\pi^2 n^2}{c^2},\;\;\;\;\kappa_{mn;ad}^2=\frac{\pi^2 m^2}{a^2}+\frac{\pi^2 n^2}{d^2}
\label{5a}
\ee

It is worth to note that $\kappa_{mn;bc}$ and $\kappa_{mn;ad}$ define the energy levels corresponding to the respective horizontal POC $H_1$ and
the vertical one $V_1$ in LSB of Fig.2 with the Dirichlet conditions on their singular diagonals while the both Fourier
expansions in \mref{1f} are nothing but the expansions of $\Psi(x,y)$ by the eigenfunctions of the corresponding POC.

Note also that the respective Fourier expansions of $\Psi(x,y)$ in POC $H_2$ or $V_2$ are determined completely by the expansions \mref{1f} so
that the latter are of the unique importance in our further considerations. This note will be valid also in the generalization of the method done
below in Sec.2.2.

The form of the quantization conditions \mref{6} suggests a reduction of the number of the independent coefficients by the following relations
\be
H_{mn}^{(1)}=(-1)^{n+n_0}\frac{(\kappa^2-\kappa_{mn_0;bc}^2)n}{(\kappa^2-\kappa_{mn;bc}^2)n_0}H_{mn_0}^{(1)},\;\;\;\;\;m,n\geq 1,\; n\neq n_0 \nn\\
V_{mn}^{(1)}=(-1)^{m_0+m}\frac{(\kappa^2-\kappa_{m_0n;ad}^2)m}{(\kappa^2-\kappa_{mn;ad}^2)m_0}V_{m_0n}^{(1)},\;\;\;\;\;m,n\geq 1,\; m\neq m_0
\label{5d}
\ee
where $n_0$ and $m_0$ have been chosen arbitrarily.

The next step is obviously the substitutions of \mref{4} into \mref{6} taking into account \mref{5d} to get direct relations between the
coefficients $H_{mn_0}^{(1)}$ and $V_{m_0n}^{(1)}$. One gets
\be
H_{mn_0}^{(1)}(\kappa^2-\kappa_{mn_0;bc}^2)=
(-1)^{m_0+n_0}\frac{4\pi}{bc^2}\frac{n_0}{m_0}\sum_{l,k\geq 1}(-1)^kk\beta_{km}\sin\ll(l\pi\frac{c}{d}\r)
\frac{\kappa^2-\kappa_{m_0l;ad}^2}{\kappa^2-\kappa_{kC;ad}^2}V_{m_0l}^{(1)}\nn\\
m\geq 1
\label{5e}
\ee
and
\be
V_{m_0n}^{(1)}(\kappa^2-\kappa_{m_0n;ad}^2)=
(-1)^{m_0+n_0}\frac{4\pi}{a^2d}\frac{ m_0}{n_0}\sum_{l,k\geq 1}(-1)^ll\alpha_{nl}\sin\ll(k\pi\frac{a}{b}\r)
\frac{\kappa^2-\kappa_{kn_0;bc}^2}{\kappa^2-\kappa_{kC;bc}^2}H_{kn_0}^{(1)}\nn\\
n\geq 1
\label{5f}
\ee
while from \mref{2a} and from \mref{5d} one has
\be
(-1)^{n+n_0}c\frac{n}{n_0}\sum_{r\geq 1}\frac{\kappa^2-\kappa_{rn_0;bc}^2}{\kappa^2-\kappa_{rn;bc}^2}\beta_{mr}H_{rn_0}^{(1)}=
(-1)^{m_0+m}a\frac{m}{m_0}\sum_{r\geq 1}\frac{\kappa^2-\kappa_{m_0r;ad}^2}{\kappa^2-\kappa_{mr;ad}^2}V_{m_0r}^{(1)}\alpha_{rn}\nn\\
m,n\geq 1
\label{5g}
\ee

The equations \mref{5e}-\mref{5g} can be rewritten shortly by
\be
\sum_{k\geq 1}\ll(\Gamma_{u,k}^H(\kappa)H_{kn_0}^{(1)}+\Gamma_{u,k}^V(\kappa)V_{m_0k}^{(1)}\r)=0,\;\;\;\;u=m,n,mn,\;m,n\geq 1
\label{8}
\ee
so that the vanishing determinant of the latter
\be
\det\ll[\Gamma_{u,k}^H(\kappa),\Gamma_{u,k}^V(\kappa)\r]=0
\label{8a}
\ee
determines the set $\{\kappa_n,\;n\geq 1\}$ of the energy spectrum of LSB.

\subsubsection{The Dirichlet boundary conditions on some sides and the Neumann ones on the others}

\hskip+2em There are no troubles with applying the method to any different boundary conditions (Dirichlet or Neumann ones) put on the sides of the
billiards. Suppose that the Neumann boundary condition is put on the vertical side with the $x$-coordinate equal to $a$. Then instead of the first
series in \mref{1f} we have to consider obviously the following one
\be
\Psi_{V_1}(x,y)=\sum_{m,n\geq 1}V_{mn}^{(1)}\sin\ll(\ll(m+\fr\r)\pi\frac{x}{a}\r)\sin\ll(n\pi\frac{y}{d}\r)
\label{6c}
\ee
while the remaining steps of the respective considerations are exactly the same. In particular the coefficients $V_{mn}^{(1)}$ are now determined
by the normal derivative of $\Psi(x,y)$ on the segment $s_H$ by
\be
V_{mn}^{(1)}\ll(\kappa^2-\frac{\pi^2m^2}{a^2}-\frac{\pi^2n^2}{d^2}\r)=
(-1)^m\frac{2}{ad}\int_0^c\frac{\p\Psi(a,y)}{\p x}\sin\ll(n\pi\frac{y}{d}\r)dy
\label{7}
\ee

\begin{figure}
\begin{center}
\includegraphics[width=7cm]{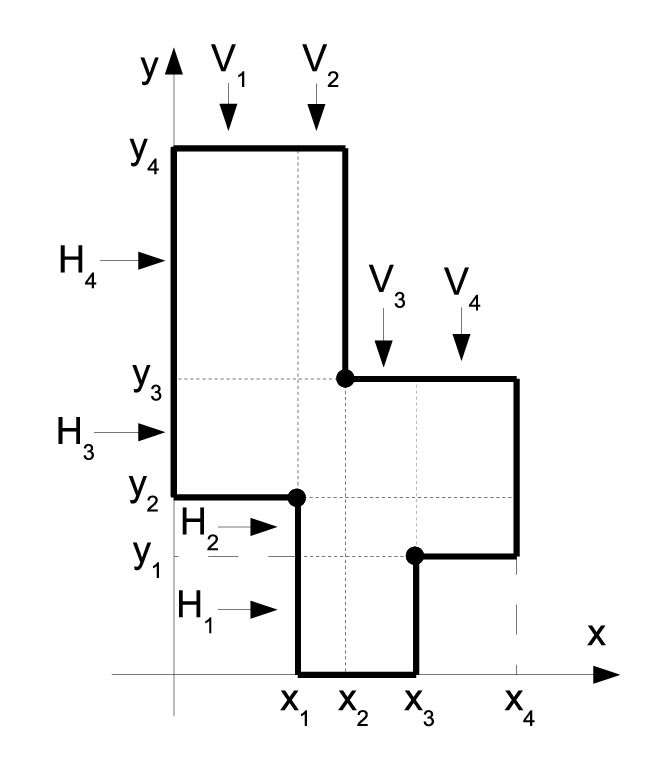}
\caption{An "arbitrary" multi-rectangular billiards. $H_1,...,H_4$ and $V_1,...,V_4$
are the horizontal and vertical POC respectively defined by the singular diagonals emerging from the corresponding vertices of the billiards}
\end{center}
\end{figure}

\subsection{Generalization to any multi-rectangular billiards}

\hskip+2em A generalization of the method to any multi-rectangular billiards is direct. Consider for example the one in Fig.2 and $\Psi(x,y)$ with the Dirichlet
boundary conditions on all its sides. Because of the fact that the respective EPP for any MRB is composed of its four images only the sine functions only can enter the
respective Fourier series expansions. Cutting the billiards respectively horizontally and vertically by the singular diagonals emerging from the
vertices distinguished in Fig.2 to form POC $H_i,V_i,\;i=1,...,4$, on it, we have the following Fourier expansions of $\Psi(x,y)$ in the
respective POC
\be
\Psi_{H_i}(x,y)=\sum_{m,n\geq 1}H_{mn}^{(i)}\sin\ll(m\pi\frac{x-x_{p_i}}{\lambda_{H_i}}\r)\sin\ll(n\pi\frac{y-y_{i-1}}{\delta_{H_i}}\r)\nn\\
\Psi_{V_i}(x,y)=\sum_{m,n\geq 1}V_{mn}^{(i)}\sin\ll(m\pi\frac{x-x_{i-1}}{\lambda_{V_i}}\r)\sin\ll(n\pi\frac{y-y_{q_i}}{\delta_{V_i}}\r)\nn\\
i=1,...,4
\label{9}
\ee
with
\be
\ba{ll}
x_{p_1}=x_{p_2}=x_1,\;x_{p_3}=x_{p_4}=0,&y_{q_1}=y_2,\;y_{q_2}=y_{q_3}=0,\;y_{q_4}=y_1\\
\lambda_{H_1}=x_3-x_1,&\delta_{H_1}=y_1\\
\lambda_{H_2}=x_4-x_1,&\delta_{H_2}=y_2-y_1\\
\lambda_{H_3}=x_4,&\delta_{H_3}=y_3-y_2\\
\lambda_{H_4}=x_2,&\delta_{H_4}=y_4-y_3\\
\lambda_{V_1}=x_1,&\delta_{V_1}=y_4-y_2\\
\lambda_{V_2}=x_2-x_1,&\delta_{V_2}=y_4\\
\lambda_{V_3}=x_3-x_2,&\delta_{V_3}=y_3\\
\lambda_{V_4}=x_4-x_3,&\delta_{V_4}=y_3-y_1
\ea
\label{10}
\ee

It is important to note that each rectangle defining a POC has to have by its construction at least on one of its horizontal sides as well as on one of its
vertical sides a piece of the multi-rectangular billiards sides because its two singular diagonals must coincides partly with some sides of the
multi-rectangle. This fact uniquely determines the form of the Fourier expansion \mref{9} for a given boundary
conditions. For the Dirichlet ones the respective form must be such as in \mref{9}.

As in the case of LSB considered the horizontal expansions have to be matched with the vertical ones in the rectangles arising
by their crossings and as in LSB case it is enough to do it as follows
\begin{itemize}
\item identify the series \mref{9} on their respective crossings;
\item write the quantization conditions for the coefficients $H_{mn}^{(i)}$ and $V_{mn}^{(i)}$;
\item express the coefficients $h_{jm}^{(i)},\;j\leq 2,$ and $v_{jn}^{(i)},\;j\leq 2,$ entering the quantization conditions by the coefficients
$H_{mn}^{(i)}$ and $V_{mn}^{(i)}$ similarly to \mref{6} - note that there are at most two coefficients $h_{jm}^{(i)},\;j\leq 2,$ and
$v_{jn}^{(i)},\;j\leq 2,$ contributing to each condition;
\end{itemize}

Denote by $a_j,\;(|a_j|=x_j-x_{j-1}),$ and $b_i,\;(|b_j|=y_i-y_{i-1}),$ respectively the horizontal and vertical sides of the rectangle $a_j\times b_i$ formed by the crossed POC
$H_i$ and $V_j$. Their lengths are equal to the respective POC halfperiods $\lambda_{V_j}$ and $\delta_{H_i}$.

Making the first step we get
\be
\delta_{H_i}\sum_{r\geq 1}\beta_{mr}^{(ji)}H_{rn}^{(i)}=\lambda_{V_j}\sum_{r\geq 1}V_{mr}^{(j)}\alpha_{rn}^{(ji)}
\label{11}
\ee
where
\be
\alpha_{rn}^{(ji)}=\int_{y_{i-1}}^{y_i}\sin\ll(r\pi\frac{y-y_{q_j}}{\delta_{V_j}}\r)\sin\ll(n\pi\frac{y-y_{i-1}}{\delta_{H_i}}\r)dyC_{ij}=\nn\\
\frac{n}{\pi\delta_{H_i}}\frac{1}{\frac{r^2}{\delta_{V_j}^2}-\frac{n^2}{\delta_{H_i}^2}}
\ll((-1)^n\sin\ll(r\pi\frac{y_j-y_{q_j}}{\delta_{V_j}}\r)-\sin\ll(r\pi\frac{y_{j-1}-y_{q_j}}{\delta_{V_j}}\r)\r)C_{ij}\nn\\
\beta_{mr}^{(ji)}=\int_{x_{j-1}}^{x_j}\sin\ll(m\pi\frac{x-x_{j-1}}{\lambda_{V_j}}\r)\sin\ll(r\pi\frac{x-x_{p_i}}{\lambda_{H_i}}\r)dxC_{ij}=\nn\\
\frac{m}{\pi\lambda_{V_j}}\frac{1}{\frac{r^2}{\lambda_{H_i}^2}-\frac{m^2}{\lambda_{V_j}^2}}
\ll((-1)^m\sin\ll(r\pi\frac{x_j-x_{p_i}}{\lambda_{H_i}}\r)-\sin\ll(r\pi\frac{x_{j-1}-x_{p_i}}{\lambda_{H_i}}\r)\r)C_{ij}
\label{11a}
\ee
where
\be
\ll[C_{ij}\r]=\left[\begin{array}{llll}
 0&1  &1  &0  \\
 0&1  &1  &1  \\
1 &1  &1  &1  \\
1 &1  &0  &0
\end{array}\right]
\ee

Realizing the second step we have
\be
H_{mn}^{(i)}\ll(\kappa^2-\ll(\kappa_{mn}^{H_i}\r)^2\r)=(-1)^n\frac{2\pi n}{\delta_{H_i}^2}(h_m^{(i,u)}-h_m^{(i,d)})\nn\\
V_{mn}^{(i)}\ll(\kappa^2-\ll(\kappa_{mn}^{V_i}\r)^2\r)=(-1)^m\frac{2\pi m}{\lambda_{V_i}^2}(v_n^{(i,r)}-v_n^{(i,l)})
\label{12}
\ee
where
\be
\ll(\kappa_{mn}^{H_i}\r)^2=\frac{\pi^2m^2}{\lambda_{H_i}^2}+\frac{\pi^2n^2}{\delta_{H_i}^2}\nn\\
\ll(\kappa_{mn}^{V_i}\r)^2=\frac{\pi^2m^2}{\lambda_{V_i}^2}+\frac{\pi^2n^2}{\delta_{V_i}^2}
\label{12f}
\ee
and
\be
h_m^{(i,u)}=\frac{2}{\lambda_{H_i}}\int_{x\in X_i}\Psi(x,y_i)\sin\ll(m\pi\frac{x-x_{p_i}}{\lambda_{H_i}}\r)dx\nn\\
h_m^{(i,d)}=\frac{2}{\lambda_{H_i}}\int_{x\in X_{i-1}}\Psi(x,y_{i-1})\sin\ll(m\pi\frac{x-x_{p_i}}{\lambda_{H_i}}\r)dx\nn\\
i=1,2,3\nn\\
X_1=[x_1,x_3],\;X_2=[x_1,x_4],\;X_3=[0,x_2]\nn\\
h_m^{(0)}=h_m^{(4)}=0
\label{12a}
\ee
so that
\be
h_m^{(i,u)}=\frac{2}{\lambda_{H_i}}\sum_{j=1}^4\delta_{jh_{ij}}\sum_{r,n\geq 1}\sin\ll(n\pi\frac{y_i-y_{q_j}}{\delta_{V_j}}\r)\beta_{rm}^{(ji)}V_{rn}^{(j)}\nn\\
h_m^{(i,d)}=\frac{2}{\lambda_{H_i}}\sum_{j=1}^4\delta_{jh_{i-1j}}\sum_{r,n\geq 1}\sin\ll(n\pi\frac{y_{i-1}-y_{q_j}}{\delta_{V_j}}\r)\beta_{rm}^{(ji)}V_{rn}^{(j)}\nn\\
i=1,2,3
\label{12b}
\ee
with the following matrix of the indeces $h_{ij}$
\be
\ll[h_{ij}\r]=\ll[\ba{llll}0&2&3&0\\0&2&3&4\\1&2&0&0\\\ea\r]
\label{12e}
\ee

The corresponding relations for the vertical POC are
\be
v_n^{(i,r)}=\frac{2}{\delta_{V_i}}\int_{y\in Y_i}\Psi(x_i,y)\sin\ll(n\pi\frac{y-y_{q_i}}{\delta_{V_i}}\r)dy\nn\\
v_n^{(i,l)}=\frac{2}{\delta_{V_i}}\int_{y\in Y_{i-1}}\Psi(x_{i-1},y)\sin\ll(n\pi\frac{y-y_{q_i}}{\delta_{V_i}}\r)dy\nn\\
i=1,2,3\nn\\
Y_1=[y_2,y_4],\;Y_2=[0,y_4],\;Y_3=[y_1,y_3]\nn\\
v_n^{(0)}=v_n^{(4)}=0
\label{12c}
\ee
so that
\be
v_n^{(i,r)}=\frac{2}{\delta_{V_i}}\sum_{j=1}^4\delta_{jv_{ij}}\sum_{s,m\geq 1}\sin\ll(m\pi\frac{x_i-x_{p_j}}{\lambda_{H_j}}\r)\alpha_{ns}^{(ij)}H_{ms}^{(j)}\nn\\
v_n^{(i,l)}=\frac{2}{\delta_{V_i}}\sum_{j=1}^4\delta_{jv_{i-1j}}\sum_{s,m\geq 1}\sin\ll(m\pi\frac{x_{i-1}-x_{p_j}}{\lambda_{H_j}}\r)\alpha_{ns}^{(ij)}H_{ms}^{(j)}\nn\\
i=1,2,3\nn\\
\label{12d}
\ee
with
\be
\ll[v_{ij}\r]=\ll[\ba{llll}0&0&3&4\\1&2&3&0\\0&2&3&0\\\ea\r]
\label{12g}
\ee

In the next step let us diminish the number of the independent amplitudes suggested by \mref{12} by
\be
H_{kl}^{(i)}=(-1)^{l+n_i}\frac{l}{n_i}\frac{\kappa^2-\ll(\kappa_{kn_i^H}^{H_i}\r)^2}{\kappa^2-\ll(\kappa_{kl}^{H_i}\r)^2}H_{kn_i^H}^{(i)}\nn\\
V_{kl}^{(i)}=(-1)^{k+m_i}\frac{k}{m_i}\frac{\kappa^2-\ll(\kappa_{m_i^Vl}^{V_i}\r)^2}{\kappa^2-\ll(\kappa_{kl}^{V_i}\r)^2}V_{m_i^Vl}^{(i)}
\label{13a}
\ee
where $m_i^V,n_i^H,\;i=1,...,4,$ are some arbitrary integers.

The next step in the procedure is the substitutions to the equations \mref{12} the respective values of $h_m^{(i)}$ and $v_n^{(i)}$ given
by \mref{12b} and \mref{12d} taking also into account \mref{13a} to get the relations between the coefficients $H_{mn_i^H}^{(i)}$ and
$V_{m_i^Vn}^{(i)}$. One gets
\be
(-1)^n\frac{4\pi n}{\lambda_{H_i}\delta_{H_i}^2}
\sum_{j=1}^4(-1)^{m_j^V}\sum_{k,l\geq 1}(-1)^k\frac{k}{m_j^V}\ll(\delta_{jh_{ij}}\sin\ll(l\pi\frac{y_i-y_{q_j}}{\delta_{V_j}}\r)-\r.\nn\\
\ll.\delta_{jh_{i-1j}}\sin\ll(l\pi\frac{y_{i-1}-y_{q_j}}{\delta_{V_j}}\r)\r)\beta_{km}^{(ji)}
\frac{\kappa^2-\ll(\kappa_{m_j^Vl}^{V_j}\r)^2}{\kappa^2-\ll(\kappa_{kl}^{V_j}\r)^2}V_{m_j^Vl}^{(j)}
=H_{mn_i^H}^{(i)}\ll(\kappa^2-\ll(\kappa_{mn_i^H}^{H_i}\r)^2\r)\nn\\
(-1)^m\frac{2\pi m}{\lambda_{V_i}^2\delta_{V_i}}
\sum_{j=1}^4(-1)^{n_j^H}\sum_{k,l\geq 1}(-1)^l\frac{l}{n_j^H}\ll(\delta_{jv_{ij}}\sin\ll(l\pi\frac{x_i-x_{p_j}}{\lambda_{H_j}}\r)-\r.\nn\\
\ll.\delta_{jv_{i-1j}}\sin\ll(l\pi\frac{x_{i-1}-x_{p_j}}{\lambda_{H_j}}\r)\r)\alpha_{nk}^{(ij)}
\frac{\kappa^2-\ll(\kappa_{ln_j^H}^{H_j}\r)^2}{\kappa^2-\ll(\kappa_{lk}^{H_j}\r)^2}H_{ln_j^H}^{(j)}
=V_{m_i^Vn}^{(i)}\ll(\kappa^2-\ll(\kappa_{m_i^Vn}^{V_i}\r)^2\r)\nn\\
i=1,...,4,\;m,n>0
\label{13}
\ee
while the relations \mref{13a} substituted to \mref{11} provide us with
\be
(-1)^{n+n_i^H}\frac{n}{n_i^H}\delta_{H_i}\sum_{r\geq 1}\beta_{mr}^{(ji)}\frac{\kappa^2-\ll(\kappa_{rn_i^H}^{H_i}\r)^2}{\kappa^2-\ll(\kappa_{rn}^{H_i}\r)^2}H_{rn_i^H}^{(i)}=\nn\\
(-1)^{m+m_j^V}\frac{m}{m_j^V}\lambda_{V_j}\sum_{r\geq 1}V_{m_j^Vr}^{(j)}\frac{\kappa^2-\ll(\kappa_{m_j^Vr}^{V_j}\r)^2}{\kappa^2-\ll(\kappa_{mr}^{V_j}\r)^2}\alpha_{rn}^{(ji)},\;\;\;\;\;
i,j=1,...,4,\;m,n>0
\label{13b}
\ee

The equations above can be rewritten shortly by
\be
\sum_{p=1}^4\sum_{r\geq 1}\ll(\Gamma_{umn,pr}^H(\kappa)H_{rn_p^H}^{(p)}+\Gamma_{umn,pr}^V(\kappa)V_{m_p^Vr}^{(p)}\r)=0\nn\\
m,n>0,\;u=ij,i,\;\;i,j=1,...,4\nn\\
\label{14}
\ee

The conditions \mref{14} define the system of the linear homogeneous equations quantizing the stationary motions
in the multi-rectangular billiards in Fig.2. The system provides us with the energy spectra determined by its vanishing determinant, i.e.
\be
\det \ll[\Gamma_{umn,pr}^H(\kappa),\Gamma_{umn,pr}^V(\kappa)\r]=0
\label{16}
\ee

\begin{figure}
\begin{center}
\includegraphics[width=7cm]{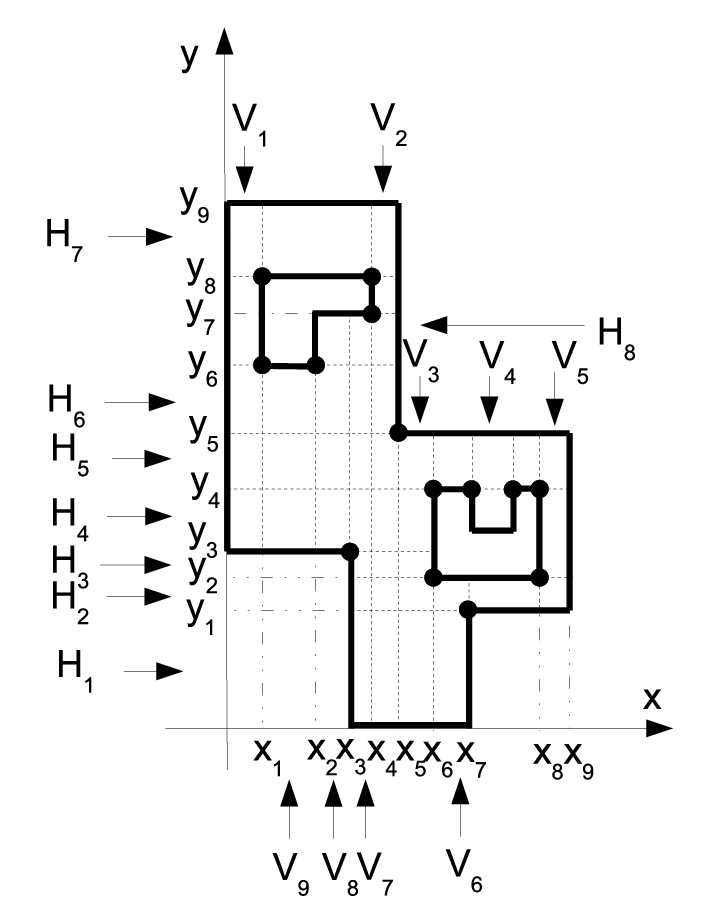}
\caption{The multi-rectangular billiards with the multi-rectangular holes. The "independent" POC are marked only.}
\end{center}
\end{figure}

It should be clear that the above method of quantization of the multi-rectangular billiards can be generalized further to the billiards with
the rectangular holes such as this shown in Fig.3, i.e. such billiards can be considered exactly in the same way -
none new element must be added to the respective procedure used in the previous examples of the multi-rectangular billiards.

\section{Superscars phenomena in the multi-rectangular billiards}

\hskip+2em The superscars phenomena mentioned first by Heller \cite{10} and next discussed widely in the context of the rational billiards by
Bogomolny and Schmit \cite{6} are still not clearly understood due to difficulties in getting of explicit formulae quantizing the rational
billiards not mentioning the chaotic ones. However the Fourier series approach to the quantization of the multi-rectangular billiards developed in the
present paper provides us with such formulae with the help of which one can try to discuss the superscars phenomena (SP) at least in the multi-rectangular
billiards, i.e. its physical meaning and conditions for their appearing.

Let us remind that by SP are understood the forms of the wave functions in the rational billiards (RB) which resembles in a part of the area of RB
the form of a (folded) wave function corresponding to some POC identified in the billiards considered. Moreover the energy level of the billiards at
which such a form appears is very close to the one of the respective POC. Bogomolny and Schmit has stressed that the superscars phenomenon is
very common for RB they considered suggesting a specific role which the energy spectra of POC can play in determining the energy spectra of RB.

\subsection{Pure POC modes in the arms of LSB}

\hskip+2em Obviously the simplest billiards of the class just considered, i.e. the rectangular one is just the one which stationary wave functions
and the energy
spectrum coincide completely with the ones of each of its two POC - the horizontal and the vertical ones whereas LSB is the
simplest one
in which such simple coincidence of the wave functions and energy spectra is impossible simply because none POC in MRB can cover it totally.

Nevertheless let us allow the possibility that in some arm of LSB say the horizontal one such a POC state can be excited to investigate respective
results of this in the second arm, i.e. assuming that in the horizontal arm the respective wave function has the form
\be
\Psi_{m_0n_0}^H(x,y)=H_{m_0n_0}^{(1)}\sin\ll(m_0\pi\frac{x}{b}\r)\sin\ll(n_0\pi\frac{y}{c}\r)
\label{17}
\ee
with the energy
\be
\kappa^2=\kappa_{m_0n_0;bc}^2=\frac{\pi^2 m_0^2}{b^2}+\frac{\pi^2 n_0^2}{c^2}
\label{18}
\ee

This assumption means therefore that $H_{kl}^{(1)}=0$ for $k,l\neq m_0,n_0$ in \mref{1f}. It then follows further from \mref{4} that
$v_n=\frac{2}{d}\alpha_{nn_0}\sin\ll(\pi m_0\frac{a}{b}\r)H_{m_0n_0}^{(1)},\;n\geq 1$, and therefore one gets from \mref{6}
\be
V_{mn}^{(1)}=(-1)^m\frac{4\pi m}{a^2d}\frac{\alpha_{nn_0}\sin\ll(\pi m_0\frac{a}{b}\r)}{\kappa_{m_0n_0;bc}^2-\kappa_{mn;ad}^2}H_{m_0n_0}^{(1)}\nn\\
m,n\geq 1
\label{18a}
\ee
which determines $\Psi_{m_0n_0}^V(x,y)$, i.e. the wave function $\Psi_{m_0n_0}(x,y)$ in the vertical arm of LSB.
But from \mref{6} one has also $h_m=0,\;m\geq 1$, so that from \mref{4} and from \mref{2a} one still gets further respectively
\be
\sin\ll(m_0\pi\frac{a}{b}\r)F_m(a,b,c,d)=0,\;\;\;\;\;m\geq 1\nn\\
\sin\ll(m_0\pi\frac{a}{b}\r)G_{mn}(a,b,c,d)=0,\;\;\;\;\;m,n\geq 1
\label{19}
\ee
where
\be
F_m(a,b,c,d)=\sum_{k,l\geq 1}\frac{(-1)^kk\beta_{km}\alpha_{ln_0}\sin\ll(l\pi\frac{c}{d}\r)}
{\kappa_{m_0n_0;bc}^2-\kappa_{kC;ad}^2},\;\;\;\;\;m\geq 1\nn\\
G_{mn}(a,b,c,d)=4\pi^2\sum_{k\geq 1}\frac{\alpha_{kn_0}\alpha_{kn}}
{\kappa_{m_0n_0;bc}^2-\kappa_{mk;ad}^2}-\frac{cd}{\frac{m_0^2}{b^2}-\frac{m^2}{a^2}}\delta_{nn_0}\nn\\
m,n\geq 1
\label{20}
\ee

Since $\sin\ll(m_0\pi\frac{a}{b}\r)\neq 0$ by the assumed irrationality of $a/b$ then \mref{19} represent an infinite number of
conditions put on the four parameters $a,b,c,d$ while the remaining
ones are numerical or are fixed like $m_0,n_0$. The conditions are functionally independent so that it seems to be impossible for any four real
number $a,b,c,d$ for which the ratio $a/b$ is irrational to satisfy them.

Nevertheless one can try to tune the LSB parameters $a,b,c,d$ making them satisfying \mref{19} at least approximately or even exactly if $a/b$
becomes rational.

\subsubsection{Rational values of $a/b$ and $c/d$ and the semiclassical solutions to LSB}

\hskip+2em Let us put therefore $a/b=p/q$ where $p,q$ are some coprime integers. Then putting $m_0=kq,\;k=1,2,...$, in \mref{19} one makes
$\sin\ll(m_0\pi\frac{a}{b}\r)$ vanishing. However in such a case the form of the second set of equations in \mref{19} is not proper for evaluating
their left hand sides and instead of them we have
\be
(-1)^m\frac{4\pi m}{acd}\sin\ll(m_0\pi\frac{a}{b}\r)\sum_{k\geq 1}\frac{\alpha_{kn_0}\alpha_{kn}}
{\kappa_{m_0n_0;bc}^2-\kappa_{mk;ad}^2}-\beta_{mm_0}\delta_{nn_0}=0\nn\\
m,n\geq 1
\label{21}
\ee
so that for $m_0=kq,\;k=1,2,...$, the l.h.s. in \mref{21} vanishes for all $m,n\geq 1,\; n\neq n_0$, but not for $n=n_0$ when the coefficients
$\beta_{kp,kq}=a/2,\;k=1,2,...$, i.e. are not equal to zero so that the respective equations in \mref{21} are not satisfied.

Nevertheless if both the ratios $a/b$ and $c/d$ are rational then there are solutions to \mref{8}. Namely if $c/d=r/s$ with two
coprime rationals $r,s$ then $\kappa_{kq,lr;bc}=\kappa_{kp,ls;cd}\equiv\kappa_{kl},\;k,l\geq 1$, define a part of the energy spectrum
corresponding to LSB with such sides and the respective stationary wave functions have the form
\be
\Psi_{kl}(x,y)=\ll\{\ba{lr}
               H_{kl}^{(1)}\sin\ll(\pi kq\frac{x}{b}\r)\sin\ll(\pi lr\frac{y}{c}\r),&(x,y)\in b\times c\\
               V_{kl}^{(1)}\sin\ll(\pi kp\frac{x}{a}\r)\sin\ll(\pi ls\frac{y}{d}\r),&(x,y)\in a\times d
               \ea\r.
\label{21a}
\ee
with $V_{kl}^{(1)}=H_{kl}^{(1)}$.

Obviously the upper row in \mref{21a} represents the POC states in the horizontal arm of LSB while the lower one the vertical POC states in LSB. Both
the states coincide in the rectangle $a\times c$ of LSB.

It should be also obvious that for LSB with the rational ratios of their sides there are also other solutions to SE different from \mref{21a}
and still determined by \mref{8} when these rational ratios of $a/b$ and $c/d$ are taken into account in them.

Let us note further that the solutions \mref{21a} coincide exactly with the ones provided by the semiclassical approximations for the $L$-shaped
billiards approximated by its doubly rational version \cite{23} with the ratios $a/b$ and $c/d$ substituted by their rational approximations $p/q$
and $r/s$ correspondingly
provided for example by their respective continued fraction representations, see \mref{26d}. In such a semiclassical approximation
the respective classical momentum is quantized taking values $p_{x,k}=\pi kq/b,\;p_{y,l}=\pi ls/d,\;k,l=\pm 1,\pm 2,...$ . Because of that the
solutions \mref{21a} corresponding to the rational ratios of $a/b=p/q$ and $c/d=r/s$ as well as the respective POC states accompanied them will
be called further semiclassical. Note that the semiclassical states of POC in LSB vanish not only on their boundaries but also on the singular
diagonals cutting them.

But even for the rational ratio of $a/b=p/q$ if $m_0\neq kq,\;k=1,2,...,$ in \mref{19}, i.e. if $m_0=kq+q_1,\;1\leq q_1\leq q-1,\;k=1,2,...,$ so that
$\sin\ll(m_0\pi\frac{a}{b}\r)=(-1)^{kp}\sin\ll(\pi\frac{q_1p}{q}\r)\neq 0$ these are the functions $F_m(a,b,c,d)=G_{mn}(a,b,c,d),\;m,n\geq 1$ which
must vanish exactly or approximately if POC \mref{17} is to be excited in LSB. However while such an expectation seems to be unrealistic it was
shown experimentally by Bogomolny {\it\underline{at} \underline{al}} \cite{18} that it can happen for the case when $q_1p/q=l+1/2,\;l=1,2,...$, which corresponds to the rectangular
billiards with a barrier inside and the respective superscar states appeared to be symmetric with respect to the barrier.

To find however that the latter possibility is
hidden in vanishing of $F_m(a,b,c,d)$ and $G_{mn}(a,b,c,d)$ even approximately is not simple since both the functions are given by the functional
series difficult to be summed to some compact forms so that only the numerical analysis of these formulae can give some hope for establishing the
presence of the superscar phenomena in them and to find necessary conditions for their appearing. Therefore in the next subsections
we will investigate only these excitations of the superscar modes which can happen when $\sin\ll(m_0\pi\frac{a}{b}\r)$ can be done close to zero
if $a/b$ is approximated by rationals.

\subsection{Closeness of energy levels of LSB to the semiclassical modes of its respective POC}

\hskip+2em Let us start with the assumption that in the spectrum $\{\kappa_n,\;n\geq 1\}$ of the energy levels of LSB there is a level
which is close to the level $\kappa_{u_0n_0;bc}$ of the $H_1$-POC. Denoting it by $\kappa_{u_0n_0}$ we can write
\be
\kappa_{u_0n_0}^2=\kappa_{u_0n_0;bc}^2+\Delta_{u_0n_0}^H
\label{22}
\ee
with
\be
|\Delta_{u_0n_0}^H|<<\frac{u_0^2}{b^2},\frac{n_0^2}{c^2}
\label{22a}
\ee
by the assumption.

Obviously assuming the last relation one also expects that in the first series in \mref{1f} the term with the coefficient $H_{u_0n_0}^{(1)}$
will dominate
the series manifesting this by the superscar effect in the horizontal arm of the billiards while the contributions of the remaining terms in both
the series in \mref{1f} to $\Psi(x,y;\kappa_{u_0n_0})$ will be clearly smaller controlled somehow by $\Delta_{u_0n_0}^H$.

Further by the assumption \mref{22} which in fact
fixes the value of energy the quantization equations \mref{8} and \mref{8a} become now the ones which should allow us to determine all the coefficients in the series
\mref{1f} by the coefficient $H_{u_0n_0}^{(1)}$ with some accuracy determined by $\Delta_{u_0n_0}^H$ defining their order of smallness with respect to $H_{m_0n_0}^{(1)}$ but
leaving the latter undefined. Therefore the homogeneous equations \mref{8} should be first transformed into the inhomogeneous ones the free parameters of
which should be determined by $H_{u_0n_0}^{(1)}$.

Consider therefore the matrix of the equations \mref{8} assuming the coefficients multiplying the amplitude $H_{u_0n_0}^{(1)}$ to form its first
column. The latter has the form
\be
\ba{cr}
&\\
0&u_0>m>0\\
\Delta_{u_0n_0}^H&m=u_0\\
0&m>u_0\\
&\\
\Delta_{u_0n_0}^H\sin\ll(u_0\pi\frac{a}{b}\r)(-1)^{u_0+n_0}\frac{4\pi u_0}{a^2dn_0}
\sum_{\ba{l}l\geq 1\\l\neq n_0\ea}\frac{(-1)^ll\alpha_{nl}}{\kappa_{u_0n_0}^2-\kappa_{u_0l;bc}^2}+\\
&\\
(-1)^{u_0}\frac{4\pi u_0}{a^2d}\alpha_{nn_0}\sin\ll(u_0\pi\frac{a}{b}\r)&n>0\\
&\\
(-1)^{m+n+n_0}\frac{cm}{\pi a}\frac{n}{n_0}\frac{\Delta_{u_0n_0}^H\sin\ll(u_0\pi\frac{a}{b}\r)}{\ll(\kappa_{u_0n_0}^2-\kappa_{u_0n;bc}^2\r)\ll(\frac{u_0^2}{b^2}-\frac{m^2}{a^2}\r)}&m>0,\;n_0>n>0\\
&\\
(-1)^m\frac{cm}{\pi a}\frac{\sin\ll(u_0\pi\frac{a}{b}\r)}{\frac{u_0^2}{b^2}-\frac{m^2}{a^2}}&m>0,\;n=n_0\\
&\\
(-1)^{m+n+n_0}\frac{cm}{\pi a}\frac{n}{n_0}\frac{\Delta_{u_0n_0}^H\sin\ll(u_0\pi\frac{a}{b}\r)}{\ll(\kappa_{u_0n_0}^2-\kappa_{u_0n;bc}^2\r)\ll(\frac{u_0^2}{b^2}-\frac{m^2}{a^2}\r)}&m>0,\;n>n_0\\
&
\ea
\label{22b}
\ee
while the remaining columns the forms
\be
\ba{ccl}
(\kappa_{u_0n_0}^2-\kappa_{kn_0;bc}^2)\delta_{mk}&\Gamma_{m,l}^V(\kappa_{u_0n_0})&k>0,\;k\neq u_0,\;l>0,\;u_0>m>0\\
0&\Gamma_{u_0,l}^V(\kappa_{u_0n_0})&k>0,\;k\neq u_0,\;l>0,\;m=u_0\\
(\kappa_{u_0n_0}^2-\kappa_{kn_0;bc}^2)\delta_{mk}&\Gamma_{m,l}^V(\kappa_{u_0n_0})&k>0,\;k\neq u_0,\;l>0,\;m>u_0\\
\Gamma_{m,k}^H(\kappa_{u_0n_0})&(\kappa_{u_0n_0}^2-\kappa_{u_0n;ad}^2)\delta_{nl}&k>0,\;k\neq u_0,\;l>0,\;n>0\\
\Gamma_{mn,k}^H(\kappa_{u_0n_0})&\Gamma_{mn,l}^V(\kappa_{u_0n_0})&k>0,\;k\neq u_0,\;l>0,\;m,n>0
\ea
\label{22c}
\ee

Therefore the equation \mref{8a} must have the form
\be
\det\ll[\Gamma_{u,k}^H(\kappa_{u_0n_0}),\Gamma_{u,k}^V(\kappa_{u_0n_0})\r]=\nn\\
A(\kappa_{u_0n_0})\Delta_{u_0n_0}^H+
B(\kappa_{u_0n_0})\Delta_{u_0n_0}^H\sin\ll(u_0\pi\frac{a}{b}\r)+C(\kappa_{u_0n_0})\sin\ll(u_0\pi\frac{a}{b}\r)=0
\label{23}
\ee
so that when $\Delta_{u_0n_0}^H=0$, i.e. when $\kappa=\kappa_{u_0n_0;bc}$ one gets
\be
\det\ll[\Gamma_{u,k}^H(\kappa_{u_0n_0;bc}),\Gamma_{u,k}^V(\kappa_{u_0n_0;bc})\r]=C(\kappa_{u_0n_0;bc})\sin\ll(u_0\pi\frac{a}{b}\r)\neq 0
\label{24}
\ee
since as it was shown earlier $\kappa_{u_0n_0;bc}$ cannot belong to the energy spectrum of LSB if $a/b$ is irrational.

Let us now invoke the assumed smallness of $\Delta_{u_0n_0}^H$ and calculate the determinant by its following
linear approximation
\be
A(\kappa_{u_0n_0;bc})\Delta_{u_0n_0}^H+
B(\kappa_{u_0n_0;bc})\Delta_{u_0n_0}^H\sin\ll(u_0\pi\frac{a}{b}\r)+C(\kappa_{u_0n_0;bc})\sin\ll(u_0\pi\frac{a}{b}\r)+\nn\\
C'(\kappa_{u_0n_0;bc})\Delta_{u_0n_0}^H\sin\ll(u_0\pi\frac{a}{b}\r)\approx 0
\label{25}
\ee
from which one gets
\be
\Delta_{u_0n_0}^H\approx-\frac{C(\kappa_{u_0n_0;bc})\sin\ll(u_0\pi\frac{a}{b}\r)}{A(\kappa_{u_0n_0;bc})+
(B(\kappa_{u_0n_0;bc})+C'(\kappa_{u_0n_0;bc}))\sin\ll(u_0\pi\frac{a}{b}\r)}
\label{26}
\ee

Note however again that if $a/b$ is a rational number, i.e. $a/b=p/q$ so that $\sin\ll(\pi kq\frac{a}{b}\r)=0,\;k=1,2,...,$ then the result
\mref{23} is completely different, i.e.
\be
\det\ll[\Gamma_{u,k}^H(\kappa_{kqn_0}),\Gamma_{u,k}^V(\kappa_{kqn_0})\r]=A_1(\kappa_{kqn_0})\Delta_{kqn_0}^H+B_1(\kappa_{kqn_0})=0
\label{26a}
\ee
and then
\be
\Delta_{kqn_0}^H\approx-\frac{B_1(\kappa_{kqn_0;bc})}{A_1(\kappa_{kqn_0;bc})}
\label{26b}
\ee

Now both the above formulae should confirm the assumed smallness of $\Delta_{u_0n_0}^H$. Looking at \mref{26} it is seen that the
obvious quantity which can determine this smallness in the irrational case of $a/b$ is the factor $\sin\ll(u_0\pi\frac{a}{b}\r)$ the argument of
which should be then close to $k\pi$ for
some integers $k$. Since $a/b$ is irrational this can happen only approximately although with any accuracy. Namely
it is well known that each real number can be approximated with an arbitrary accuracy by a rational one so this can be done also with the ratio $a/b$.
It is therefore clear that $u_0$ in $\sin\ll(u_0\pi\frac{a}{b}\r)$ can be chosen in such a way to make $u_0a/b$ close to an integer with any accuracy.
To realize this one can use for example the continued fraction representations for $a/b$ which can approximate it by the fractions $p_n/q_n,\;n\geq 1$,
with the accuracy better than $1/q_n^2$, i.e.
\be
\ll|\frac{a}{b}-\frac{p_n}{q_n}\r|<\frac{1}{q_n^2}
\label{26d}
\ee
where $q_n,p_n\to\infty$ if $n\to\infty$.

Obviously putting $u_0=kq_n,\;k=1,2,...$, for each such an approximation one gets
\be
\ll|\sin\ll(\pi u_0\frac{a}{b}\r)\r|=\ll|\sin\ll(\pi kq_n\ll(\frac{a}{b}-\frac{p_n}{q_n}\r)\r)\r|<\frac{k\pi}{q_n}
\label{27a}
\ee
i.e. taking $q_n$ sufficiently large and choosing $k<<q_n,\;k=1,2,...$, one can make $\Delta_{kq_n,n_0}^H$ arbitrarily small.

The inequality \mref{27a} means also that the
$H_1$-POC state $H_{kq_nn_0}^{(1)}\sin\ll(\pi kq_n\frac{x}{b}\r)\sin\ll(\pi n_0\frac{y}{c}\r)$ vanishes approximately on the singular diagonal $s_V$
of LSB of Fig.2, i.e. it is approximately semiclassical POC state.

Note further that in the rational case of $a/b$ there is no obvious reason for the formula \mref{26b} to provide us with the small
$\Delta_{u_0n_0}^H$.

Therefore assuming the amplitude $H_{kq_nn_0}$ to be known one could next solve the equations \mref{8} with respect to the remaining amplitudes
concluding that they are of the order of $k/q_n$ smaller than $H_{kq_nn_0}$, i.e. that the contribution of the latter to the series \mref{1f} is
dominating. Unfortunately it is not true since if $u_0=kq_n$ then the coefficient $\beta_{mkq_n}$ in \mref{5g} for $m=kp_n$ does not produce the
small factor $\sin\ll(\pi kq_n\frac{a}{b}\r)$ but up to $1/q_n^2$ it becomes equal to $a/2$ which means that in the column \mref{22b} its element with
$m=kp_n,n=n_0$ equal to $c\beta_{kp_nkq_n}$ is not small so that \mref{23} changes to
\be
A(\kappa_{kq_nn_0})\Delta_{kq_nn_0}^H+
B(\kappa_{kq_nn_0})\Delta_{kq_nn_0}^H\sin\ll(kq_n\pi\frac{a}{b}\r)+C(\kappa_{kq_nn_0})\sin\ll(kq_n\pi\frac{a}{b}\r)+\nn\\
D(\kappa_{kq_nn_0})c\beta_{kp_nkq_n}=0
\label{28}
\ee
and consequently $\Delta_{kq_nn_0}^H$ calculated from \mref{28} cannot be small.

It is easy to note that this negative result corresponds to the one provided by the equations \mref{21} which also cannot be satisfied by the same
reason, i.e one therefore can conclude that one cannot excite a semiclassical POC state in the horizontal arm of LSB only even approximately.

It is seen however that it is the equation \mref{2a} with $m=kp_n,n=n_0$ which is the source of the troubles in which the coefficient $\beta_{kp_nkq_n}$
is not small and the term $\beta_{kp_nkq_n}H_{kq_nn_0}^{(1)}$ must be compensated somehow by some term on the r.h.s. of the equality \mref{2a}. One can
easily guess that for this one should try to excite some POC in the vertical arm  of LSB which would coincide approximately with the $H_1$-one in the
rectangle $a\times c$. A respective POC state can be constructed in the way similar to the one in the $H_1$-arm approximating the
quotient $c/d$ by respective rationals $r_n/s_n,\;n>0$ analogously to the $a/b$ one. Doing
this one can then choose in the equations \mref{5e}-\mref{5g} $m_0=kp_n,\;n_0=lr_n$ and then the coefficient $\alpha_{ls_n,lr_n}$ in the equation
\mref{5g} up to the order $1/s_n^2$ becomes equal to $c/2$. Now one notes that since $\kappa_{u_0n_0}\equiv\kappa_{kq_nlr_n}$ we have
\be
\ll|\Delta_{kq_nlr_n}^H-\Delta_{kp_nls_n}^V\r|=\ll|\kappa_{kq_nlr_n;bc}^2-\kappa_{kp_nls_n;ad}^2\r|<
\frac{k^2}{q_n^2}\frac{3b}{a}\frac{q_n^2}{b^2}+\frac{l^2}{s_n^2}\frac{3d}{c}\frac{s_n^2}{d^2}<<\frac{q_n^2}{b^2}+\frac{s_n^2}{d^2}
\label{29}
\ee
i.e. the level $\kappa_{kq_nlr_n}$ is also close to the level $\kappa_{kp_nls_n;ad}$ of the $V_1$-POC of LSB. Therefore we
rename it putting further $\kappa_{kq_nlr_n}\equiv\kappa_{kp_nls_n}\equiv\kappa_{kl}^{(n)}$.

We can write therefore
\be
\Delta_{kq_nlr_n}^H=\Delta_{kl}^{(n)}+O_H\ll(\frac{k^2}{q_n^2},\frac{l^2}{s_n^2}\r)\nn\\
\Delta_{kp_nls_n}^V=\Delta_{kl}^{(n)}+O_V\ll(\frac{k^2}{q_n^2},\frac{l^2}{s_n^2}\r)
\label{30}
\ee

Therefore forming now analogously to \mref{22b} the first two columns of the determinant \mref{8a} by the coefficients multiplying the amplitudes
$H_{kq_nlr_n}^{(1)}$ and $V_{kp_nls_n}^{(1)}$ respectively one gets for the first one
\be
\ba{ll}
&\\
0&kq_n>m>0\\
\Delta_{kq_nlr_n}^H&m=kq_n\\
0&m>kq_n\\
&\\
\Delta_{kq_nlr_n}^H\sin\ll(kq_n\pi\frac{a}{b}\r)(-1)^{kq_n+lr_n}\frac{4\pi kp_n}{a^2dlr_n}\times\\
&\\
\sum_{\ba{l}u\geq 1\\u\neq lr_n\ea}\frac{(-1)^uu\alpha_{vu}}{\ll(\kappa_{kl}^{(n)}\r)^2-\kappa_{kq_nu;bc}^2}+
(-1)^{kq_n}\frac{4\pi kq_n}{a^2d}\alpha_{vlr_n}\sin\ll(kq_n\pi\frac{a}{b}\r)&v>0\\
&\\
(-1)^{v+lr_n}\frac{cv}{lr_n}\frac{\beta_{kp_nkq_n}}{\ll(\kappa_{kl}^{(n)}\r)^2-\kappa_{kq_nv;bc}^2}\Delta_{kq_nlr_n}^H&m=kp_n,\;v>0,\\
&v\neq lr_n\\
&\\
c\beta_{mkq_n}&m>0,\;m\neq kp_n,\\
&v=lr_n\\
&\\
c\beta_{kp_nkq_n}&m=kp_n,\;v=lr_n\\
&
\ea
\label{28a}
\ee
and for the second one
\be
\ba{ll}
&\\
\Delta_{kp_nls_n}^V\sin\ll(ls_n\pi\frac{c}{d}\r)(-1)^{kp_n+lr_n}\frac{4\pi lr_n}{c^2bkp_n}\times\\
&\\
\sum_{\ba{l}u\geq 1\\u\neq lr_n\ea}\frac{(-1)^uu\beta_{um}}{\ll(\kappa_{kl}^{(n)}\r)^2-\kappa_{kq_nu;bc}^2}+
(-1)^{lr_n}\frac{4\pi lr_n}{bc^2}\beta_{kp_nlr_n}\sin\ll(ls_n\pi\frac{c}{d}\r)&m>0\\
&\\
0&kp_n>v>0\\
\Delta_{kp_nls_n}^V&v=kp_n\\
0&v>kp_n\\
&\\
(-1)^{m+kp_n}\frac{cm}{kp_n}\frac{\alpha_{ls_nv}}{\ll(\kappa_{kl}^{(n)}\r)^2-\kappa_{mls_n;ad}^2}\Delta_{kp_nls_n}^V&m>0,\;m\neq kp_n\\
&v>0,\;v\neq lr_n\\
&\\
a\alpha_{ls_nv}&m=kp_n\\
&v>0,n\neq lr_n\\
&\\
(-1)^{m+kp_n}\frac{cm}{kp_n}\frac{\alpha_{ls_nlr_n}}{\ll(\kappa_{kl}^{(n)}\r)^2-\kappa_{mls_n;ad}^2}\Delta_{kp_nls_n}^V&m>0,\;m\neq kp_n\\
&v=lr_n\\
&\\
a\alpha_{ls_nlr_n}&m=kp_n,\;v=lr_n\\
&
\ea
\label{28b}
\ee
so that for the the determinant \mref{8a} one gets up to the order $1/q_n^2,1/s_n^2$
\be
\tilde{A}_n\Delta_{kl}^{(n)}+\tilde{B}_n\sin\ll(\pi kq_n\frac{a}{b}\r)+\tilde{C}_n\sin\ll(\pi ls_n\frac{c}{d}\r)=0
\label{31}
\ee
which proves the smallness of $\Delta_{kl}^{(n)}$.

Therefore assuming that apart from the amplitude $H_{kq_nlr_n}$ also the amplitude $V_{kp_nls_n}$ is known one can solve the equations \mref{5e}-\mref{5f}
with respect to the remaining amplitudes concluding their following property
\be
H_{mlr_n}^{(1)}=A_m^H\sin\ll(\pi kq_n\frac{a}{b}\r)H_{kq_nlr_n}^{(1)}+B_m^H\sin\ll(\pi ls_n\frac{c}{d}\r)V_{kp_nls_n}^{(1)}\nn\\
V_{kp_nv}^{(1)}=A_v^V\sin\ll(\pi kq_n\frac{a}{b}\r)H_{kq_nlr_n}^{(1)}+B_v^V\sin\ll(\pi ls_n\frac{c}{d}\r)V_{kp_nls_n}^{(1)}\nn\\
\;\;\;\;\;\;\;m,v\geq 1,\;m\neq kq_n,\;v\neq ls_n
\label{32f}
\ee
and by the equations \mref{5d}
\be
H_{mv}^{(1)}=A_{mv}^H\sin\ll(\pi kq_n\frac{a}{b}\r)H_{kq_nlr_n}^{(1)}+B_{mv}^H\sin\ll(\pi ls_n\frac{c}{d}\r)V_{kp_nls_n}^{(1)}\nn\\
V_{mv}^{(1)}=A_{mv}^V\sin\ll(\pi kq_n\frac{a}{b}\r)H_{kq_nlr_n}^{(1)}+B_{mv}^V\sin\ll(\pi ls_n\frac{c}{d}\r)V_{kp_nls_n}^{(1)}\nn\\
\;\;\;\;\;\;\;m,v\geq 1,\;(m,v)\neq(kq_n,ls_n)
\label{33d}
\ee
as well.

Substituting the above results to the equation \mref{5g} with the indeces $m=m_0=kp_n,\;v=n_0=lr_n$ and solving it with respect to the amplitudes
$H_{kq_nlr_n}^{(1)},\;V_{kp_nls_n}^{(1)}$ one gets
\be
H_{kq_n,lr_n}^{(1)}=V_{kp_n,ls_n}^{(1)}+O(k/q_n,l/s_n)
\label{31f}
\ee
as well as
\be
\ll|\sin\ll(\pi kq_n\frac{x}{b}\r)\sin\ll(\pi lr_n\frac{y}{c}\r)-\sin\ll(\pi kp_n\frac{x}{a}\r)\sin\ll(\pi ls_n\frac{y}{d}\r)\r|
<2\pi\ll(\frac{bk}{aq_n}+\frac{dl}{cs_n}\r)
\label{31d}
\ee
i.e. if $k/q_n,\;l/s_n<<1$ then in both the series \mref{1f} the dominating terms are those which correspond to the semiclassical modes
\mref{21a} of the $L$-shaped
billiards with the rational ratios of their sides, i.e. $a/b=p_n/q_n$ and $c/d=r_n/s_n$ approximating the one considered.

Note now that the result \mref{26b} for the rational $a/b=p/q$ is now changed by making $B_1(\kappa_{kqn_0;bc})$ to be proportional to
$\sin\ll(\pi n_0\frac{c}{d}\r)$ which makes $\Delta_{kqn_0}^H$ to be small by the latter factor if $c/d$ is approximated by rationals. Note
further that if $c/d=r/s$, i.e. is also rational then $\Delta_{kqn_0}^H$ for $n_0=lr$ vanishes since $\kappa_{kqlr}$ becomes then equal to
$\kappa_{kqlr;bc}=\kappa_{kpls;ad}$.

The results obtained up to now can be done much more transparent in the following way. Let us assume the rational approximations \mref{26d} and the
respective ones for the ratio $c/d$ and consider the $L_n$-shaped billiards with the sides $a,b_n,c,d_n$ where
\be
b_n=\frac{q_n}{p_n}a,\;\;\;d_n=\frac{s_n}{r_n}c
\label{31e}
\ee
so that
\be
|b-b_n|<\frac{b}{p_nq_n},\;\;\;|d-d_n|<\frac{d}{r_ns_n}
\label{31g}
\ee

This $L_n$-shaped billiards is the doubly rational approximation of the $L$-one. Its exact solutions \mref{21a} with $b=b_n,\;d=d_n$
coincide with their semiclassical approximations \cite{23}.

Let us note now that the $L_n$-shaped billiards can be transformed smoothly into the $L$-one by the following transformation
\be
\ba{lr}
x'=x,\;y'=y,\;\;\;&(x,y)\in [0,a]\times [0,c]\\
x'=\frac{b_n-a}{b-a}x+\frac{b-b_n}{b-a}a,\;y'=y,\;\;\;&(x,y)\in [a,b]\times [0,c]\\
x'=x,\;y'=\frac{d_n-c}{d-c}y+\frac{d-d_n}{d-c}c,\;\;\;&(x,y)\in [0,a]\times [c,d]
\ea
\label{32a}
\ee
with
\be
|x-x'|<\frac{b}{p_nq_n},\;\;|y-y'|<\frac{d}{r_ns_n},\;\;(x,y)\in L
\label{32b}
\ee

According to the Theorem 1 of App.B if $\{\kappa_m,\;m\geq 1\}$ is energy spectrum of LSB while $\{\kappa_{n,m},\;m\geq 1\}$
of the $L_n$-one then we have
\be
\ll|\frac{\kappa_{n,m}}{\kappa_m}-1\r|<\eta_n
\label{32c}
\ee
with $\eta_n\to 0$ if $n\to\infty$.

In particular the results got in the last section show that
\begin{itemize}
\item for sufficiently large $n$ the respective partners in the energy spectrum of LSB for the semiclassical levels
$\kappa_{kq_nlr_n;b_nc}=\kappa_{kp_nls_n;ad_n}$ of the $L_n$-one satisfying \mref{32c} are
$\kappa_{kl}^{(n)}=\kappa_{kq_nlr_n;b_nc}+{\Delta}_{kl}^{(n)}$ since ${\Delta}_{kl}^{(n)}=\Delta_{kq_nlr_n}^H+
\omega_{kl}^H$ where
\be
|\omega_{kl}^H|=|\kappa_{kq_nlr_n;b_nc}-\kappa_{kq_nlr_n;bc}|<\frac{3k^2}{p_nq_n}\frac{q_n^2}{b^2}<<\frac{q_n^2}{b^2}
\label{32d}
\ee
if $k<<\sqrt{p_nq_n}$, i.e. $\omega_{kl}^H$ is of the same order of smallness as $\Delta_{kl}^{(n)}$, see \mref{22}; and moreover
\item the series \mref{1f} defining the wave function $\Psi(x,y;\kappa_{kl}^{(n)})$ are then dominated in the respective arms of LSB by their
single terms
\[H_{kl}^{(1)}\sin\ll(\pi kq_n\frac{x}{b}\r)\sin\ll(\pi lr_n\frac{y}{c}\r)\] and \[V_{kl}^{(1)}
\sin\ll(\pi kp_n\frac{x}{a}\r)\sin\ll(\pi ls_n\frac{y}{d}\r)\] which according to \mref{31f}-\mref{31d} coincide approximately in the rectangle
$a\times c$ of LSB, i.e.
\be
H_{kl}^{(1)}\sin\ll(\pi kq_n\frac{x}{b}\r)\sin\ll(\pi lr_n\frac{y}{c}\r)\approx V_{kl}^{(1)}
\sin\ll(\pi kp_n\frac{x}{a}\r)\sin\ll(\pi ls_n\frac{y}{d}\r)\nn\\
(x,y)\in a\times c
\label{32e}
\ee
\end{itemize}

Therefore the important conclusions which follows from the last results are
\begin{itemize}
\item one can excite in LSB the semiclassical modes \mref{21a} of the respective $L_n$-shaped
billiards which approximate the original one;
\item one cannot excite a semiclassical POC mode of an $L_n$-shaped billiards in some arm of LSB not exciting
immediately the respective one in the second arm;
\item the excited semiclassical states in LSB are collective semiclassical superscars states of both the POC in LSB - the horizontal and the
vertical one;
\item one can excite in the original LSB an infinite number of the semiclassical super scars states which energies are close to
the semiclassical
ones of the respective $L_n$-shaped billiards approximating the $L$-one by increasing the accuracy of the approximations \mref{26d} for $a/b$ and
the respective for $c/d$, i.e. by increasing $n$;
\item the higher energetically are modes of LSB approximated semiclassically, the stronger is the effect of the semiclassical
super scars exciting;
\item the paper of Kudrolli and Sridhar \cite{20} confirms the above conclusions at the experimental level.
\end{itemize}

\subsection{The superscar phenomena in the multi-rectangular billiards}

\subsubsection{DRMRB approximating MRB}

\hskip+2em In the previous subsection the conditions have been formulated for the superscar states to be excited in LSB
with the conclusion that these superscar states corresponds to the semiclassical states of the respective doubly rational $L$-shape billiards
approximating the original one.
It is clear that the respective discussion can be easily extended to the general case of the multi-rectangular billiards following the methods
used in the case of LSB to get similar results. To show this we will consider as "general" the case of MRB shown in Fig.3
approximating it by the doubly rational MRB (DRMRB) shown in Fig.5 in the following way.
\begin{figure}
\begin{center}
\includegraphics[width=7cm]{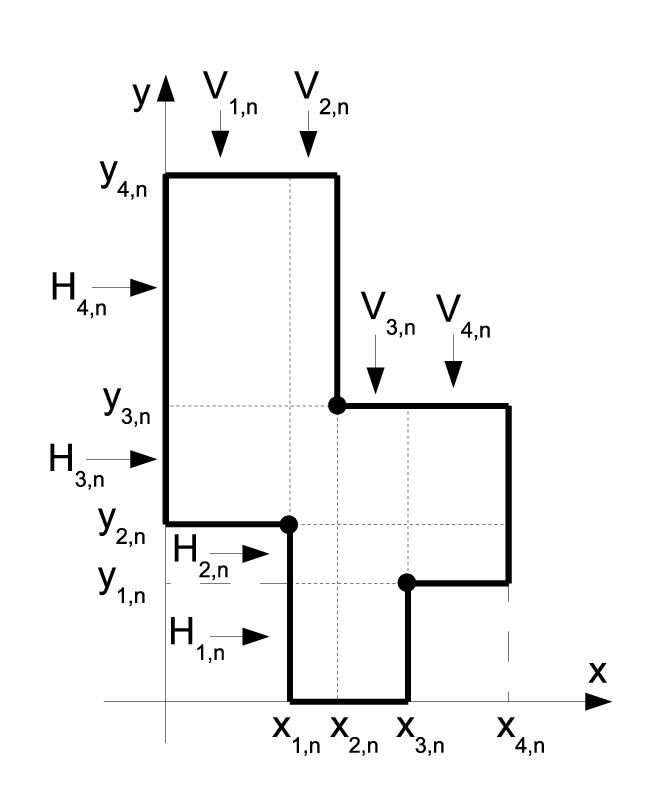}
\caption{DRMRB approximating MRB of Fig.3}
\end{center}
\end{figure}

Let us first approximate the respective ratios of the coordinates $x_i,y_i$ using the respective theorem of Dirichlet of App.C. For the case
considered it takes the form
\be
\ll|\frac{x_i}{x_1}-\frac{p_{n,i}}{q_{n,i}}\r|<\frac{1}{C_{n,1}\sqrt[3]{n}},\;\;\;
\ll|\frac{y_i}{y_1}-\frac{r_{n,i}}{s_{n,i}}\r|<\frac{1}{D_{n,1}\sqrt[3]{n}},\nn\\
i=2,3,4,\;\;\;C_{n,1},\;D_{n,1}\leq n
\label{32}
\ee
where $C_{n,1}$ is the least common multiple (LCM) of $q_{n,i}$ while $D_{n,1}$ of $s_{n,i},\;i=1,2,3$ and $n$ is an arbitrary integer.

Let us now define DRMRB approximating MRB considered by the following set of its coordinates $x_{n,i},\;y_{n,i},\;i=1,...,4$
\be
x_{n,1}=x_1,\;x_{n,i}=\frac{p_{n,i}}{q_{n,i}}x_1\nn\\
y_{n,1}=y_1,\;y_{n,i}=\frac{r_{n,i}}{s_{n,i}}y_1\nn\\
i=2,3,4
\label{33}
\ee
so that
\be
|x_i-x_{n,i}|<\frac{x_1}{C_{n,1}\sqrt[3]{n}},\;\;\;|y_i-y_{n,i}|<\frac{y_1}{D_{n,1}\sqrt[3]{n}},\;\;\;i=1,...,4
\label{33a}
\ee

Denote $a_i=[x_i,x_{i-1}],\;b_i=[y_i,y_{i-1}],\;i=1,2,3,4$, with $x_0=0,\;y_0=0$. Then the approximated MRB can be continuously transformed into
the constructed DRMRB by
\be
\ba{lr}
(x,y)\to(x',y')=(X_i(x),Y_j(y)),\;\;\;(x,y)\in a_i\times b_j&\\
X_i=x+\frac{x-x_i}{x_i-x_{i-1}}(x_{i-1}-x_{n,i-1})-\frac{x-x_{i-1}}{x_i-x_{i-1}}(x_i-x_{n,i})&\\
Y_j=y+\frac{y-y_j}{y_j-y_{j-1}}(y_{j-1}-y_{n,j-1})-\frac{y-y_{j-1}}{y_j-y_{j-1}}(y_j-y_{n,j})&\\
i,j=1,2,3,4,\;\;(i,j)\neq(1,1),(1,2),(3,4),(4,1),(4,4)&
\ea
\label{33b}
\ee
with $x_{n,0}=y_{n,0}=0$ so that
\be
|x'-x|<\frac{2x_1}{\lambda}\frac{1}{q_n^2},\;|y'-y|<\frac{2y_1}{\delta}\frac{1}{s_n^2}
\label{33c}
\ee
where $\lambda$ is minimal of the periods $\lambda_{V_i},\;i=2,3,4$ while $\delta$ - of the periods $\delta_{H_i},\;i=2,3,4$ and
$q_n,s_n$ denote the minimal numbers in the sets $\{q_{n,i},\;i=2,3,4\}$ and $\{s_{n,i},\;i=2,3,4\}$ respectively.

Let us further put $C_{n,i}=u_ip_{n,i}$ if $C_{n,1}=u_iq_{n,i}$ and $D_{n,i}=w_ir_{n,i}$ if $D_{n,1}=w_is_{n,i}$, and let yet $\delta_{H_{n,i}}$ and $\lambda_{V_{n,j}}$ be given by \mref{10} where
$x_i,y_i$ are substituted by $x_{n,i},y_{n,i},\;i=2,3,4$. Then
\be
\frac{C_{n,i}-C_{n,j}}{x_{n,i}-x_{n,j}}=\frac{C_{n,1}}{x_1},\;\;\;\;\frac{D_{n,i}-D_{n,j}}{y_{n,i}-y_{n,j}}=\frac{D_{n,1}}{y_1}\nn\\
i,j=1,...,4
\label{34}
\ee
Let $C_{n,H_i}=C_{n,u_i}-C_{n,w_i}$ if $\lambda_{H_{n,i}}=x_{n,u_i}-x_{n,w_i},\;i=1,2,3,4$ and respectively $D_{n,V_i}=D_{n,u_i}-D_{n,w_i}$ if
$\delta_{V_{n,i}}=y_{n,u_i}-y_{n,w_i},\;i=1,2,3,4$. Then the wave functions
\be
\Psi_{kl}^{(n)}(x,y)=\nn\\
\ll\{\ba{r}(-1)^{(kC_{n,w_i}+lD_{n,i-1})}A_{kl}\sin\ll(\pi kC_{n,H_i}\frac{x-x_{n,w_i}}{\lambda_{H_{n,i}}}\r)
\sin\ll(\pi l(D_{n,i}-D_{n,i-1})\frac{y-y_{n,i-1}}{\delta_{H_{n,i}}}\r),\\
\\
(x,y)\in[x_{n,w_i},x_{n,u_i}]\times[y_{n,i-1},y_{n,i}]\\
\\
(-1)^{(kC_{n,i-1}+lD_{n,w_i})}A_{kl}\sin\ll(\pi k(C_{n,i}-C_{n,i-1})\frac{x-x_{n,i-1}}{\lambda_{V_{n,i}}}\r)
\sin\ll(\pi lD_{n,V_i}\frac{y-y_{n,w_i}}{\delta_{V_{n,i}}}\r),\\
\\
(x,y)\in[x_{n,i-1},x_{n,i}]\times[y_{n,w_i},y_{n,u_i}]
\ea\r.\nn\\
\nn\\
i=1,...,4
\label{35}
\ee
are the semiclassical solutions for the approximating DRMRB.

Note again that the solutions in the first row of \mref{35} are the semiclassical states in the horizontal POC $H_{i,n},\;i=1,...,4$ in DRMRB
while in the second row the the semiclassical states in the respective vertical ones $V_{i,n},\;i=1,...,4$, i.e. the semiclassical solutions
\mref{35} are collected from the semiclassical states of the horizontal and vertical POC in DRMRB coinciding in the rectangles
$[x_{n,j-1},x_{n,j}]\times[y_{n,i-1},y_{n,i}]$
formed by the crossed POC $H_{i,n}$ and $V_{j,n},\;i,j=1,...,4$, i.e. the semiclassical states of POC vanish on its boundary as well as on
each singular diagonal which crosses POC.

The quantized semiclassical momenta corresponding to the states \mref{35} are \[p_{x,k}^{(n)}=\pi kC_{n,1}/x_1,\;p_{y,l}^{(n)}=
\pi lD_{n,1}/y_1,\;k,l=1,2,...\].

Taking into account the relations \mref{33c} and Theorem 1 of App.B we can claim that between the energy spectrum $\kappa_m,\;m\geq 1$, of MRB
of Fig.3 and
the one $\kappa_m^{(n)},\;m\geq 1$, of its DRMRB approximation defined by \mref{33b} there is one to one correspondence for which the relation \mref{32c}
is satisfied. In particular to the energies $(\kappa_{kl}^{(n)})^2=k^2C_{n,1}^2/x_1^2+l^2D_{n,1}^2/y_1^2,\;k,l\geq 1,$ of the states \mref{35} there
correspond the states in MRB with energies $\kappa_{kl}^2=(\kappa_{kl}^{(n)})^2+\Delta_{kl}^{(n)}$ which satisfy the condition \mref{32c}, i.e.
\be
\ll|\frac{\ll(\kappa_{kl}^{(n)}\r)^2}{\kappa_{kl}^2}-1\r|<\eta_n
\label{36}
\ee
so that $\Delta_{kl}^{(n)}<<(\kappa_{kl}^{(n)})^2$ for sufficiently large $n$.

The arising question whether the terms
\be
H_{kl}^{(i)}\sin\ll(\pi kC_{n,H_i}\frac{x-x_{p_i}}{\lambda_{H_i}}\r)\sin\ll(\pi l(D_{n,i}-D_{n,i-1})\frac{y-y_{i-1}}{\delta_{H_i}}\r)\nn\\
V_{kl}^{(i)}\sin\ll(\pi k(C_{n,i}-C_{n,i-1})\frac{x-x_{i-1}}{\lambda_{V_i}}\r)\sin\ll(\pi lD_{n,V_i}\frac{y-y_{q_i}}{\delta_{V_i}}\r)\nn\\
i=1,...,4
\label{37}
\ee
dominate the series \mref{9} can be answered in the same way as in the case of LSB by solving the quantization conditions
\mref{14} while assuming the form $\kappa_{kl}^2=(\kappa_{kl}^{(n)})^2+\Delta_{kl}^{(n)}$ for the respective part of the energy spectrum.
The main trouble in the respective procedure is the number of the eight amplitudes $H_{kl}^{(i)},V_{kl}^{(i)}\;i=1,...,4$, which have
to be taken into account in the corresponding calculations. The procedure is described in the next section.

\subsubsection{The excitations of the semiclassical modes in the multi-rectangular billiards}

\hskip+2em Consider the semiclassical level $(\kappa_{kl}^{(n)})^2=k^2C_{n,1}^2/x_1^2+l^2D_1^2/y_1^2,$ of DRMRB of Fig.5 approximating the
original one of Fig.3. Then according to Theorems 1. and 2. of App.B there is the energy level $\kappa_{kl}^2=(\kappa_{kl}^{(n)})^2+\Delta_{kl}^{(n)}$
of MRB of Fig.2 which together with  $(\kappa_{kl}^{(n)})^2$ satisfy \mref{36} so that
\be
\Delta_{kl}^{(n)}<<k^2C_{n,1}^2/x_1^2+l^2D_1^2/y_1^2
\label{38}
\ee
for sufficiently large $n$.

Putting next in \mref{13} and \mref{13b} $m_s^V=k(C_s-C_{s-1},\;n_s^H=l(D_s-D_{s-1}),\;s=1,2,3,4$ one gets from \mref{11} and \mref{13} the following quantization
conditions for the considered MRB
\be
\ba{ll}
{\bf(A)}&\ll\{\ba{l}
\sum_{j=1}^4\sum_{k,l\geq 1}\ll(\delta_{jh_{ij}}\sin\ll(l\pi\frac{y_i-y_{q_j}}{\delta_{V_j}}\r)-
\delta_{jh_{i-1j}}\sin\ll(l\pi\frac{y_{i-1}-y_{q_j}}{\delta_{V_j}}\r)\r)\times\\
(-1)^{k+m_j^V}\frac{k}{m_j^V}\frac{\kappa_{kcl}^2-\ll(\kappa_{m_j^Vl}^{V_j}\r)^2}{\kappa_{kl}^2-\ll(\kappa_{kl}^{V_j}\r)^2}\beta_{km}^{(ji)}V_{m_j^Vl}^{(j)}
-\\
\frac{(-1)^{n_i^H}\delta_{H_i}^2\lambda_{H_i}}{4\pi n_i^H}\ll(\kappa_{kl}^2-\ll(\kappa_{mn_i^H}^{H_i}\r)^2\r)H_{mn_i^H}^{(i)}=0\\
\\
\sum_{j=1}^4\sum_{k,l\geq 1}\ll(\delta_{jv_{ij}}\sin\ll(k\pi\frac{x_i-x_{p_j}}{\lambda_{H_j}}\r)-
\delta_{jv_{i-1j}}\sin\ll(k\pi\frac{x_{i-1}-x_{p_j}}{\lambda_{H_j}}\r)\r)\times\\
(-1)^{l+n_j^H}\frac{l}{n_j^H}\alpha_{nl}^{(ij)}\frac{\kappa_{kl}^2-\ll(\kappa_{kn_j^H}^{H_j}\r)^2}{\kappa_{kl}^2-\ll(\kappa_{kl}^{H_j}\r)^2}H_{kn_j^H}^{(j)}-
\\
\frac{(-1)^{m_i^V}\lambda_{V_i}^2\delta_{V_i}}{4\pi m_i^V}{\ll(\kappa_{kl}^2-\ll(\kappa_{m_i^Vn}^{V_i}\r)^2\r)}V_{m_i^Vn}^{(i)}=0\\
i=1,...,4,\;m,n>0
\ea\r.\\
&\\
{(\bf B)}&\ll\{\ba{l}
(-1)^{n+n_i^H}\frac{n}{n_i^H}\delta_{H_i}\sum_{k\geq 1}\beta_{mk}^{(ji)}
\frac{\kappa_{kl}^2-\ll(\kappa_{kn_i^H}^{H_i}\r)^2}{\kappa_{kl}^2-\ll(\kappa_{kn}^{H_i}\r)^2}H_{kn_i^H}^{(i)}-\\
(-1)^{m+m_j^V}\frac{m}{m_j^V}\delta_{V_j}\sum_{k\geq 1}\alpha_{kn_i^H}^{(ji)}
\frac{\kappa_{kl}^2-\ll(\kappa_{m_j^Vk}^{V_j}\r)^2}{\kappa_{kl}^2-\ll(\kappa_{mk}^{V_j}\r)^2}V_{m_j^Vk}^{(j)}=0\\
i,j=1,...,4,\;m,n>0
\ea\r.
\ea
\label{39}
\ee

Now let us solve the first group {\bf(A)} of the above equations considering all the amplitudes $H_{m_i^Hn_i^H}^{(i)}$ and $V_{m_j^Vn_j^V}^{(j)}$
with $m_i^H=kC_{n,H_i},\;n_j^V=lD_{n,V_j}$ as known. Taking into account the following relations
\be
\kappa_{kl}^2-\ll(\kappa_{m_i^Hn_i^H}^{H_i}\r)^2=\Delta_{kl}^{(n)}+\delta_{kl}^{H_i}\nn\\
\kappa_{kl}^2-\ll(\kappa_{m_i^Vn_i^V}^{V_i}\r)^2=\Delta_{kl}^{(n)}+\delta_{kl}^{V_i}
\label{40}
\ee
one has
\be
\ll|\delta_{kl}^{H_i}\r|=\ll|\kappa_{kl}^{(n)})^2-\ll(\kappa_{m_i^Hn_i^H}^{H_i}\r)^2\r|<\frac{3}{\lambda_{H_i}^2}\frac{x_1}{C_{n,1}\sqrt[3]{n}}\frac{k^2C_{n,1}^2}{x_1^2}+
                        \frac{3}{\delta_{H_i}^2}\frac{y_1}{D_{n,1}\sqrt[3]{n}}\frac{l^2D_{n,1}^2}{y_1^2}\nn\\
\ll|\delta_{kl}^{V_i}\r|=\ll|\kappa_{kl}^{(n)})^2-\ll(\kappa_{m_i^Vn_i^V}^{V_i}\r)^2\r|<\frac{3}{\lambda_{V_i}^2}\frac{x_1}{C_{n,1}\sqrt[3]{n}}\frac{k^2C_{n,1}^2}{x_1^2}+
                        \frac{3}{\delta_{V_i}^2}\frac{y_1}{D_{n,1}\sqrt[3]{n}}\frac{l^2D_{n,1}^2}{y_1^2}
\label{41}
\ee
i.e. both $\delta_{kl}^{H_i}$ and $\delta_{kl}^{V_i}$ are of the same order as $\Delta_{kl}^{(n)}$.

Next due to the following inequalities
\be
\ll|\sin\ll(\pi k(C_{n,c}-C_{n,d})\frac{x_a-x_b}{x_c-x_d}\r)\r|<2\pi\ll(1+\ll|\frac{x_a-x_b}{x_c-x_d}\r|\r)\frac{k}{\sqrt[3]{n}}\nn\\
\ll|\sin\ll(\pi l(D_{n,c}-D_{n,d})\frac{y_a-y_b}{y_c-y_d}\r)\r|<2\pi\ll(1+\ll|\frac{y_a-y_b}{y_c-y_d}\r|\r)\frac{l}{\sqrt[3]{n}}\nn\\
\label{42}
\ee
one can see that the coefficients in the group {\bf (A)} of the equations \mref{39} multiplying the amplitudes $H_{m_i^Hn_i^H}^{(i)}$ and $V_{m_j^Vn_j^V}^{(j)}$
are of the order $n^{-1/3}$ and are small for sufficiently large $n$ and $k,l<<n^{-1/3}$.

Therefore taking also into account the relations \mref{13a} one can find the coefficients $H_{mn}^{(i)}$ and $V_{mn}^{(i)}$ with $mn$ different
than $m_i^Hn_i^H$ and $m_i^Vn_i^V$ respectively to be smaller by the factor $n^{-1/3}$ than the amplitudes $H_{m_i^Hn_i^H}^{(i)}$ and $V_{m_j^Vn_j^V}^{(j)}$.
Taking this result into account in the group {\bf (B)} of the equations \mref{39} one gets
\be
(-1)^{k(C_{n,j-1}-C_{n,p_i})}H_{m_i^Hn_i^H}^{(i)}C_{ij}=(-1)^{l(D_{n,i-1}-D_{n,q_j})}V_{m_j^Vn_j^V}^{(j)}C_{ij}+O_{ij}(n^{-1/3})
\label{43}
\ee

As the final conclusions one gets therefore the following ones
\begin{itemize}
\item there are infinitely many energy levels $\kappa_{kl}^2$ of MRB close to the semiclassical ones of the corresponding DRMRB;
\item if such levels are excited their respective wave functions represented by the series \mref{9} are dominated by the amplitudes
$H_{m_i^Hn_i^H}^{(i)}$ and $V_{m_j^Vn_j^V}^{(j)},\;i=1,...,4,$ with respect to the other ones by the factor $n^{1/3}$ and the respective states
resonate with the semiclassical ones \mref{35} of DRMRB approximating the original billiards. The resonant effects are the stronger the higher
energies are excited.
\end{itemize}

\subsection{Degenerated MRB - MRB with barriers}

\hskip+2em The results of the investigations done above remain valid also for the degenerated MRB such as in Fig.7. But as we have mentioned earlier
\begin{figure}
\begin{center}
\includegraphics[width=7cm]{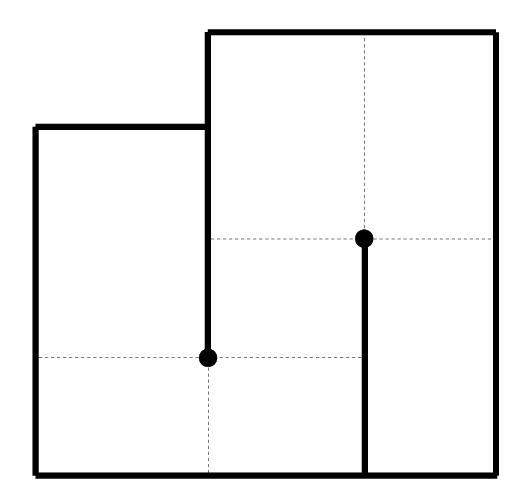}
\caption{The degenerated MRB - MRB with the barriers}
\end{center}
\end{figure}
the quantization conditions \mref{14}-\mref{16} which certainly covered the superscars effects observed and discussed for
such billiards by Bogomolny {\it\underline{at} \underline{al}} \cite{18,16} seem to be ineffective for a discussing the latter by the methods applied above for
the semiclassical excitations, i.e. the respective analyses wait to be discovered.

\section{Summary and conclusions}

\hskip+2em In this paper the Fourier series expansions have been used to quantize the multi-rectangu-\linebreak lar billiards, i.e. the rational billiards which
each
angle is $\pi/2$ or $3\pi/2$. The method has appeared to be very
effective allowing to get in compact forms the wave functions as well as the corresponding conditions for the energy spectra which allowed for
further analysis of them looking for the superscar states. The main result of this analysis which has been found is
\begin{enumerate}
\item the superscars states which can be excited in MRB are of two kinds
\begin{itemize}
\item the semiclassical ones corresponding to the semiclassical states of MRB which approximate the original one being doubly
rational, i.e. the ratios between their horizontal side as well as between their vertical ones are rational - such DRMRB can approximate the original
one with any accuracy; and
\item the remaining ones corresponding to the non-semiclassical states of the horizontal and vertical POC covering MRB not investigated in the present
paper but found in the respective experiment \cite{18};
\end{itemize}
\end{enumerate}

While the resonating semiclassical states of the DRMRB approximating MRB considered are not a surprise
since such a possibility follow directly from the general theorems ruling the subject (see App.B) there are the following new things
concerning the latter relations and shown in the paper
\begin{enumerate}
\setcounter{enumi}{1}
\item the resonating states of DRMRB dominate the respective ones of the original MRB;
\item there are infinitely many of states of the original MRB dominated by the corresponding states of DRMRB approximating the former;
\item the resonating states of DRMRB are their semiclassical ones;
\item since every semiclassical state of DRMRB is simultaneously an eigenstate of each of the horizontal and the vertical POC of DRMRB then one can
consider the semiclassical superscars states in MRB also as an effect of a collective resonating of POC mentioned which is observed in the
experiment \cite{20}.
\end{enumerate}

Finally one has to conclude also that the quantization conditions established by the Fourier series approach to the problem seem to be ineffective
for the respective investigations the superscars states found and discussed by Bogomolny {\it\underline{at} \underline{al}} \cite{18,16}.

\appendix

\section{The Fourier series expansion in the rectangle \cite{13}}

\hskip+2em Suppose a function $\Psi(x,y)$ is given in the rectangle $a\times b$, see Fig.5, inside which it is of the class $C^2$ with respect to
its both variables. This function can always be extended into the one defined in the three other rectangles in the figure by the following conditions
\begin{figure}
\begin{center}
\includegraphics[width=7cm]{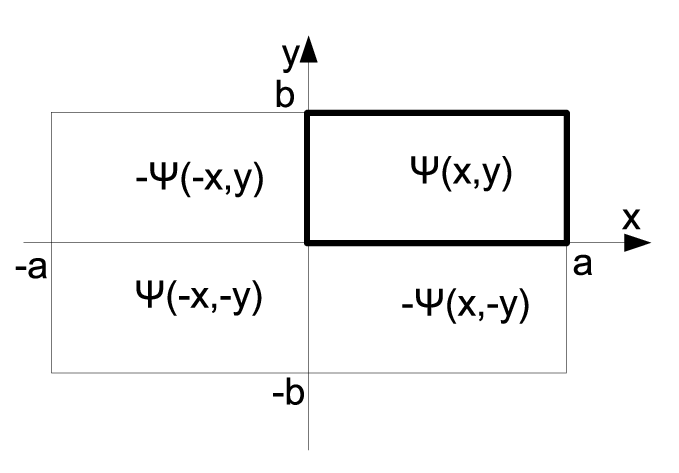}
\caption{The rectangular billiards $a\times b$ in which $\Psi(x,y)$ is defined and next extended antysymmetrically to $\Phi(x,y)$ defined in
the rectangle with the sides $2a\times 2b$ (see App.A)}
\end{center}
\end{figure}
\be
\Phi(x,y)=\Psi(x,y),\;\;0<x<a,\;0<y<b\nn\\
\Phi(-x,y)=\Phi(x,-y)=-\Phi(-x,-y)=-\Phi(x,y),\;\;0<x<a,\;0<y<b\nn\\
\Phi(0,y)=\Phi(-a,y)=\Phi(a,y)=0,\;\;\;-b\leq y\leq b\nn\\
\Phi(x,0)=\Phi(x,-b)=\Phi(x,b)=0,\;\;\;-a\leq x\leq a
\label{A4}
\ee
i.e. $\Phi(x,y)$ is the antisymmetric function of its variable in the rectangle $2a\times 2b$. In this rectangle the function can be discontinues with the
properties
\be
\Phi(0_{\pm},y)=\pm\Psi(0_+,y),\;\;\;0\leq y\leq b\nn\\
\Phi(x,0_{\pm})=\pm\Psi(x,0_+),\;\;\;0\leq x\leq a \nn\\
\Phi(\pm a_{\mp},y)={\pm}\Psi(a_-,y),\;\;\;0\leq y\leq b\nn\\
\Phi(x,\pm b_{\mp})={\pm}\Psi(x,b_-),\;\;\;0\leq x\leq b
\label{A4a}
\ee
and can be expanded into the following Fourier series
\be
\Phi^{FS}(x,y)=\sum_{m,n\geq 1}X_{mn}\sin\ll(m\pi\frac{x}{a}\r)\sin\ll(n\pi\frac{y}{b}\r)
\label{A5}
\ee
with the property
\be
\Phi(x,y)=\Phi^{FS}(x,y),\;\;\;-a\leq x\leq a,\;-b\leq y\leq b
\label{A6}
\ee
while
\be
\Psi(x,y)=\Phi^{FS}(x,y),\;\;\;0< x< a,\;0< y< b
\label{A6a}
\ee
i.e. $\Psi(x,y)$ itself is reconstructed by the series \mref{A5} inside the rectangle $a\times b$.

Obviously the above series makes $\Phi(x,y)$ extended periodically on the whole plane.

\begin{figure}
\begin{center}
\includegraphics[width=7cm]{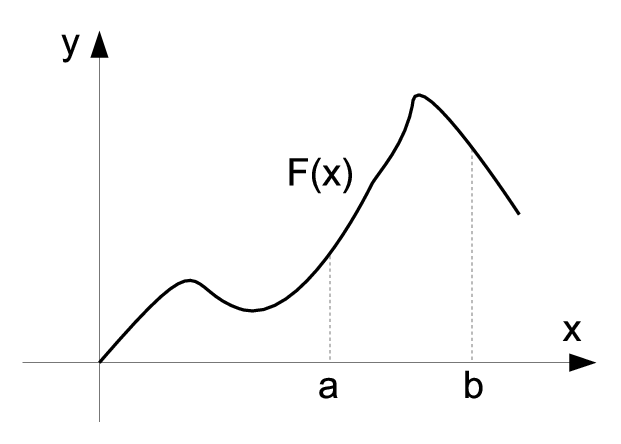}
\caption{A function $F(x)$ the Fourier series properties of which are discussed in App.A}
\end{center}
\end{figure}

The coefficients $X_{mn}$ are given by
\be
X_{mn}=\frac{1}{ab}\int_{-a}^adx\int_{-b}^bdy\Phi(x,y)\sin\ll(m\pi\frac{x}{a}\r)\sin\ll(n\pi\frac{y}{b}\r)=\nn\\
\frac{4}{ab}\int_0^adx\int_0^bdy\Psi(x,y)\sin\ll(m\pi\frac{x}{a}\r)\sin\ll(n\pi\frac{y}{b}\r),\;\;\;m,n\geq 1
\label{A7}
\ee

The derivatives of $\Psi(x,y)$ can be also extended into the rectangle $2a\times 2b$ by the conditions analogous to \mref{A4} with the respective
Fourier series. Namely, for the $x$-derivatives we have
\be
\ll(\frac{\p \Phi(x,y)}{\p x}\r)^{FS}=\sum_{m,n\geq 1}X_{mn}^{(x)}\cos\ll(m\pi\frac{x}{a}\r)\sin\ll(n\pi\frac{y}{b}\r)\nn\\
\ll(\frac{\p^2 \Phi(x,y)}{\p x^2}\r)^{FS}=\sum_{m,n\geq 1}X_{mn}^{(x^2)}\sin\ll(m\pi\frac{x}{a}\r)\sin\ll(n\pi\frac{y}{b}\r)
\label{A8}
\ee
with
\be
X_{mn}^{(x)}=\frac{1}{ab}\int_{-a}^adx\int_{-b}^bdy\frac{\p \Phi(x,y)}{\p x}\cos\ll(m\pi\frac{x}{a}\r)\sin\ll(n\pi\frac{y}{b}\r)=\nn\\
\frac{4}{ab}\int_0^adx\int_0^bdy\frac{\p \Psi(x,y)}{\p x}\cos\ll(m\pi\frac{x}{a}\r)\sin\ll(n\pi\frac{y}{b}\r)=\nn\\
\frac{4}{ab}\int_0^bdy\ll((-1)^m\Psi(a,y)-\Psi(0,y)\r)\sin\ll(n\pi\frac{y}{b}\r)+\frac{m\pi}{a}X_{mn}\nn\\
X_{mn}^{(x^2)}=\frac{4}{ab}\int_0^adx\int_0^bdy\frac{\p^2 \Psi(x,y)}{\p x^2}\sin\ll(m\pi\frac{x}{a}\r)\sin\ll(n\pi\frac{y}{b}\r)=\nn\\
-\frac{4\pi m}{a^2b}\int_0^bdy\ll((-1)^m\Psi(a,y)-\Psi(0,y)\r)\sin\ll(n\pi\frac{y}{b}\r)-\frac{m^2\pi^2}{a^2}X_{mn},\;\;\;m,n\geq 1
\label{A9}
\ee
where the integration by parts and the Green theorem have been applied.

Similar formulae can be got for the remaining derivatives.

Consider further a function $F(x)$ given on the line for simplicity, see Fig.8, and its two Fourier series expansions in the two
different segments - $[0,a]$ and $[0,b]$ ones. They can be the following
\be
\Phi_a(x)=\sum_{m\geq 1}A_m\sin\ll(\pi m\frac{x}{a}\r)\nn\\
\Phi_b(x)=\sum_{m\geq 1}B_m\sin\ll(\pi m\frac{x}{b}\r)
\label{A17}
\ee
with the coefficients given by
\be
A_m=\frac{2}{a}\int_{0}^aF(x)\sin\ll(\pi m\frac{x}{a}\r)dx\nn\\
B_m=\frac{2}{b}\int_{0}^bF(x)\sin\ll(\pi m\frac{x}{b}\r)dx
\label{A18}
\ee
i.e. we have chosen the antisymmetric extension $\Phi(x)$ of $F(x)$ into the negative segments $[-a,0]$ and $[-b,0]$. Due to that we have
\be
\Phi_a(a)=\fr(F(a)+(-F(a))=0\nn\\
\Phi_b(b)=\fr(F(b)+(-F(b))=0
\label{A19}
\ee

The respective Fourier expansions for the first and second derivatives of $F(x)$ are
\be
\Phi'_a(x)=\sum_{m\geq 1}A'_m\cos\ll(\pi m\frac{x}{a}\r)\nn\\
\Phi'_b(x)=\sum_{m\geq 1}B'_m\cos\ll(\pi m\frac{x}{b}\r)\nn\\
\Phi''_a(x)=\sum_{m\geq 1}A''_m\sin\ll(\pi m\frac{x}{a}\r)\nn\\
\Phi''_b(x)=\sum_{m\geq 1}B''_m\sin\ll(\pi m\frac{x}{b}\r)
\label{A20}
\ee
while their Fourier coefficients are related to the ones of $F(x)$ by
\be
A'_m=(-1)^mF(a)+\frac{\pi m}{a}A_m\nn\\
B'_m=(-1)^mF(b)+\frac{\pi m}{b}B_m\nn\\
A''_m=(-1)^{m+1}\frac{\pi m}{a}F(a)-\frac{\pi^2 m^2}{a^2}A_m\nn\\
B''_m=(-1)^{m+1}\frac{\pi m}{b}F(b)-\frac{\pi^2 m^2}{b^2}B_m
\label{A21}
\ee

The series \mref{A17} and \mref{A20} must coincide on the segment $[0,a)$ so that we must have
\be
A_m=\frac{2}{a}\sum_{n\geq 1}\alpha_{mn}B_n\nn\\
A'_m=\frac{2}{a}\sum_{n\geq 1}\frac{an}{bm}\alpha_{mn}B'_n\nn\\
A''_m=\frac{2}{a}\sum_{n\geq 1}\alpha_{mn}B''_n
\label{A22}
\ee
where
\be
\alpha_{mn}=\int_0^a\sin\ll(\pi m\frac{x}{a}\r)\sin\ll(\pi n\frac{x}{b}\r)=
\frac{(-1)^mm}{\pi a}\frac{\sin\ll(\pi n\frac{a}{b}\r)}{\frac{n^2}{b^2}-\frac{m^2}{a^2}},\;\;\;\;m,n\geq 1
\label{A23}
\ee

However because of the relations \mref{A21} the second group and the third one of the equations \mref{A22} can be reduced to the first group of them,
i.e. the latter group of the equations \mref{A22} is sufficient to form the necessary and sufficient conditions for the respective coincidence of $F(x)$ with itself on the segment $[0,a)$.
This important result is then used further in the main body of the paper.

\section{Smooth behavior of energy levels as a function of a billiard boundary - general theorems \cite{13}}

\hskip+2em Consider two billiards which are close to each other in the meaning of the following theorem proved in the monograph of Courant and Hilbert.

\begin{de} It is said that the domain $G$ is approximated by the domain $G'$ with the $\epsilon$-accuracy if $G$ together with its
boundary can be transformed pointwise into the domain $G'$ together with its boundary by the equations
\begin{eqnarray}
x'=x+g(x,y)\nonumber\\
y'=y+h(x,y)
\label{A10}
\end{eqnarray}
where $g(x,y),\;h(x,y)$ are both piecewise continuous and less in $G$ in their absolute values than a small positive number $\epsilon$ together with
their first derivatives.
\end{de}

\begin{de}
If all conditions of Definition 1 are satisfied while $\epsilon\to 0$ then it is said that $G$ is a continuous deformation of $G'$.
\end{de}

\begin{tw}
Let $G$ and $G'$ satisfy all conditions of Definition 1. Then for any boundary condition $\partial\Psi/\partial n+\sigma\Psi=0$ the energy spectrum
corresponding to $G'$ approximates the one of $G$ with the $\epsilon$-accuracy. More precisely for any $\epsilon$ there is a number $\eta$
depending only on $\epsilon$ and vanishing with it such that for respectively ordered energy levels $E'_n$ and $E_n$ corresponding to the domains
$G'$ and $G$ we have
\begin{eqnarray}
\left|\frac{E_n'}{E_n}-1\right|<\eta
\label{A11}
\end{eqnarray}
\end{tw}

\begin{tw}
Let $G$ and $G'$  satisfy the conditions of Theorem 1  and $G$ is a continuous deformation of $G'$ then the energy spectrum
corresponding to $G'$ varies continuously with $\epsilon\to 0$ approaching the energy spectrum of $G$ controlled by the conditions (\ref{A11}).
\end{tw}

\begin{tw}
Theorem 1 remains valid with none condition on the first derivatives of $g(x,y),\;h(x,y)$ in the case of the Dirichlet boundary condition $\Psi=0$.
\end{tw}

\begin{tw}
If $G$ and $G'$ are transformed each into other by (\ref{A10}) and the absolute value of the Jacobean of the latter transformation is bounded from above
and below than the ratio $E_n'/E_n$ for respectively ordered energy levels $E'_n$ and $E_n$ corresponding to the domains $G'$ and $G$ satisfy for
sufficiently large $n$ the following relation
\begin{eqnarray}
0<a<\left|\frac{E_n'}{E_n}\right|<b
\label{A12}
\end{eqnarray}
where $a$ and $b$ are independent of $n$.
\end{tw}

\section{The Dirichlet simultaneous approximation theorem (DAT) \cite{14}}

\begin{tw}
For any real numbers $X_1,...,X_n$ and any natural $N$ there exist integers $K_1,...,K_n$ and $0<C_X\leq N$ which satisfy the condition
\begin{eqnarray}
|C_XX_i-K_i|<\frac{1}{N^\frac{1}{n}},\;\;\;\;\;\;\;   0<C_X\leq N\nonumber\\
i=1,...,n
\label{A13}
\end{eqnarray}
\end{tw}

Let us call each set \{$K_1/C_X,...,K_n/C_X$\} of rationals satisfying for some $N$ the conditions (\ref{A7}) a rationalization of the set
\{$X_1,...,X_n$\}.

It can happen that the real numbers $X_1,...,X_n$ can be linearly expressed by a smaller number of other real numbers $Y_1,...,Y_m,\;m<n$ with rational
coefficients, i.e.
\begin{equation}
X_i=\sum_{j=1}^m\frac{p_{ij}}{q_{ij}}Y_j,\;\;\;\;\;\;i=1,...,n
\label{A14}
\end{equation}

In such a case one can first rationalize by DAT the set \{$Y_1,...,Y_m,\;m<n$\} and next rationalize the set \{$X_1,...,X_n$\} by putting in (\ref{A7})
$C_X=CC_Y$ where $C$ is the least common multiple of all the denominators $q_{ij}$ in (\ref{A8}), i.e. $C=n_{ij}q_{ij}$ with integer $n_{ij}$. This
allows us to improve the approximations (\ref{A7}) by increasing the exponent of $N$ from $1/n$ to $1/m$. Namely we have
\begin{equation}
|CC_YX_i-\sum_{j=1}^mn_{ij}p_{ij}H_j|<\frac{\sum_{j=1}^mn_{ij}|p_{ij}|}{N^\frac{1}{m}},\;\;\;\;\;\;i=1,...,n
\label{A15}
\end{equation}
if
\begin{equation}
|C_YY_i-H_i|<\frac{1}{N^\frac{1}{m}},\;\;\;\;\;\;i=1,...,m
\label{A16}
\end{equation}

\end{document}